\documentclass[onecolumn,floatfix,10pt,nofootinbib]{revtex4}
\pdfoutput=1

\usepackage[utf8]{inputenc}
\usepackage{amsfonts}
\usepackage{amssymb}
\usepackage{amsmath}
\usepackage{boxedminipage}
\usepackage{listings}
\usepackage{minitoc}
\usepackage{ifpdf}
\usepackage{color}

\ifpdf
\usepackage[pdftex]{graphicx}
\else
\usepackage{graphicx}
\fi

\newcommand{\ket}[1]{\left\vert#1 \right\rangle}

\definecolor{red}{rgb}{1,0,0}
\begin{document}

\title{\textbf{\large A protected vortex exciton qubit}}

\author{Suvabrata De}\email{py09sd@leeds.ac.uk}
\affiliation{School of Physics and Astronomy, University of Leeds, Leeds, LS2 9JT, United Kingdom}
\author{Tim Spiller}
\affiliation{School of Physics and Astronomy, University of Leeds, Leeds, LS2 9JT, United Kingdom}

\date{\today}

\begin{abstract}
We propose a protected qubit which is `dual' to a suggestion of a superconducting current mirror qubit [A. Kitaev, arXiv:0609441 (2006)]. Our construction can be regarded as the magnetic analogue of Kitaev's proposal: it inherits the intrinsic fault-tolerance of the current mirror qubit, but may perform better than it in the laboratory, since magnetic noise is generally less of a problem than electric noise. We adapt the scheme for universal fault-tolerant quantum computation proposed by Kitaev to our construction.   
\end{abstract}

\maketitle

\section{Introduction} \label{Intro}

The difficulty of realizing even a modest sized quantum computer stems from the ever present effects of decoherence, and the inevitable errors which are present when attempting to control quantum systems. Despite this, research into quantum fault-tolerance has taught us that we may be able to overcome these problems. One solution is to employ `software' error correction, where errors are actively detected and corrected during the computation \cite{Shor,N+C}.  An alternative method is to find robust or `topologically protected' quantum hardware which is effective in suppressing errors \cite{Kitaev}. Hybrids of these two approaches, such as surface codes \cite{Bravyi+Kitaev,Fowler}, are also conceivable, with recent experiments involving superconducting qubits reaching the surface code threshold \cite{Barends}.         

Topological protection has been sought in superconducting circuits for some time \cite{IoffeNature,Ioffe}, although these particular examples would be difficult to realize with present day technology. Encouragingly though, intrinsic fault tolerance using arrays of Josephson junctions has recently been demonstrated in the laboratory \cite{Gladchenko}. The qubit in \cite{Gladchenko} was found to be protected against local flux noises well beyond linear order, while conventional approaches, which rely on tuning the qubit control parameters, offer only linear order protection \cite{Vion,Chiorescu,Wallraff}.   

In a similar vein, Kitaev proposed a protected qubit composed of two chains of Josephson junctions which are capacitively coupled to each other \cite{KitaevJJ}. While the system consists of many microscopic degrees of freedom, information can be encoded in a global phase degree of freedom. Since the logical states can be identified with nearly degenerate ground states localised near phases of zero or $\pi$, the system belongs to a family of superconducting `$0$-$\pi$ qubits' \cite{Ioffe,Gladchenko,Doucot+Vidal}. The splitting of the degeneracy vanishes exponentially as the size of the system increases, and has been shown to survive in the presence of modest disorder in circuit parameters \cite{Dempster}. Kitaev also proposed a scheme for universal fault-tolerant quantum computation, with protected single and two-qubit phase gates being noteworthy. These phase gates have been shown to be robust against Hamiltonian and thermal noise, with the gate errors being exponentially small given a large enough $LC$ oscillator impedance \cite{Brooks}. A realization of Kitaev's qubit in a bilayer exciton condensate contacted by superconducting leads has been suggested \cite{Peotta}.   

In this paper, we construct a protected qubit analogous to Kitaev's qubit and mimic the proposed scheme for universal fault-tolerant quantum computation. In our construction, we utilise vortices and their conjugate phases as the fundamental microscopic degrees of freedom, instead of Cooper pairs and superconducting phases, which are used in Kitaev's qubit. A motivation for searching for a `dual' Kitaev qubit is that noise may be less of an issue for such a qubit. This is because there are `magnetic systems' and substrates which suffer less from flux noise, compared to the problems that `electric systems' and substrates face from charge noise. We note that quantum coherent behaviour of vortices has been observed in arrays of Josephson junctions \cite{Elion}.  

The paper is organised as follows. In Sec.~\ref{KCMQ}, we explain the emergence of a qubit in the system examined by Kitaev. Sec.~\ref{VEQ} deals with the emergence of a `dual' Kitaev qubit in a suitable array of Josephson junctions. In Sec.~\ref{M+G}, the means to perform measurements and a universal set of gates on such qubits are outlined. We conclude in Sec.~\ref{Conclusions}, while the dual basis measurement protocol in Kitaev's qubit and the construction of a vortex oscillator and a vortex DC SQUID (which are needed for implementing gates on the dual qubit) are left to the appendices.      

\begin{figure}[t]
\begin{minipage}{\columnwidth}
\begin{center}
\resizebox{.5\columnwidth}{!}{\rotatebox{0}{\includegraphics{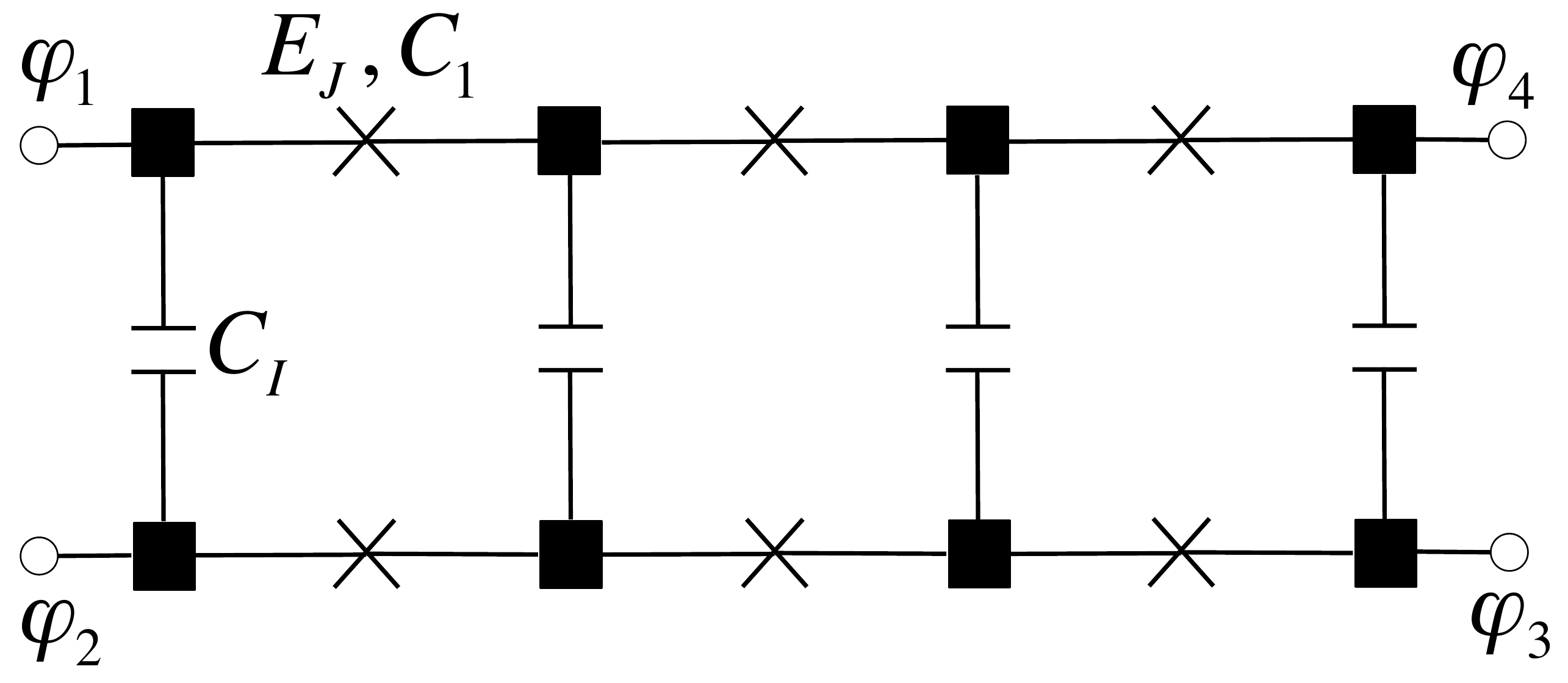}}}
\end{center}
\vspace*{-0.5cm} \caption{Schematic diagram of the superconducting current mirror, which consists of two capacitively coupled chains of Josephson junctions. The superconducting islands or grains are denoted by black squares, while the Josephson junctions are denoted by crosses. Each junction has a Josephson energy $E_J$ and a capacitance $C_1$, while the chains are coupled by a capacitance $C_I$.  The phases of the four corner grains are $\varphi_i$, and superconducting leads are attached to these grains.} \label{KitaevQubit}
\end{minipage}
\end{figure} 

\section{Kitaev's current mirror qubit} \label{KCMQ}

Kitaev's idea builds on the work of Choi {\it{et al.}} \cite{Choi}, where the authors investigated the quantum phase transitions that can take place in a system consisting of two capacitively coupled chains of Josephson junctions (see Fig.~\ref{KitaevQubit}). In this system, each of the junctions have a Josephson energy $E_J$ \footnote{We use the notation of Choi {\it{et al.}} in \cite{Choi}, which is different from Kitaev's notation.} and a charging energy $E_1 \equiv e^2/2C_1$, which is associated with the junction capacitance $C_1$. There is also another characteristic energy, $E_I \equiv e^2/2C_I$, which corresponds to the inter-chain capacitance $C_I$ which couples the chains (no tunnelling - Cooper-pair or otherwise - is allowed between the chains). In \cite{Choi}, a self-capacitance is also present, which along with the application of an external gate voltage to each superconducting island results in an external (polarization) charge being induced. However, such a finite external charge would push the system into phases which would be unsuitable for realizing a qubit, and so in Kitaev's treatment, the self-capacitance is absent. 

$C_1$ is assumed to be extremely small (i.e. $E_J <<E_1$). As a result, without inter-chain coupling, each junction and hence each chain, would be in the insulating phase: current, in the form of Cooper-pairs, cannot flow in an individual chain. It is also assumed that the coupling capacitance is much larger than the junction capacitance: $C_I>>C_1$, which means that $E_I<<E_1$. While uncorrelated individual currents cannot flow, it is energetically much more favourable for Cooper-pairs to propagate in one chain, while Cooper-pairs propagate simultaneously in the other chain in the opposite direction. This is the reason for the name `current mirror' present in the title of \cite{KitaevJJ}. In other words, excitons - Cooper-pair and `Cooper-hole' pairs - can propagate along the array. 

The Hamiltonian of the system is

\begin{equation}
\mathcal{H} = \frac{(2e)^2}{2}\sum_{l,l';x,x'} n_l(x) \mathcal{C}_{ll'}^{-1}(x,x') n_{l'}(x') -E_J \sum_{l,x} \cos \left[ \varphi_l(x+1) - \varphi_l(x) \right],
\label{H_KCMQ}
\end{equation}
where the conjugate variables $n_l(x)$ and $\varphi_l(x)$ satisfy $\left[ \varphi_l(x), n_m(y) \right] = i\delta_{lm}\delta_{xy}$, and  refer to the excess number of Cooper-pairs on, and the phase of, the island at position $x$ and on chain $l$ ($l=1,2$), respectively. The first term in (\ref{H_KCMQ}) captures the charging energy of the array, while the second term reflects the Josephson energies of the junctions. The capacitance matrix, $\mathcal{C}$, encapsulates the electrostatic interactions between charges, and is defined by

\begin{equation}
\mathcal{C}_{ll'}(x,x') \equiv C(x,x') \otimes \delta_{ll'} + C_I \delta_{xx'} \otimes \begin{pmatrix} 1 & -1 \cr -1 & 1 \cr \end{pmatrix}, 
\end{equation}
where the intra-chain capacitance matrix is

\begin{equation}
C(x,x') \equiv C_1 \left( 2\delta_{xx'} - \delta_{x,x'+1} - \delta_{x,x'-1} \right).
\end{equation}

\begin{figure}[t]
\begin{minipage}{\columnwidth}
\begin{center}
\resizebox{.5\columnwidth}{!}{\rotatebox{0}{\includegraphics{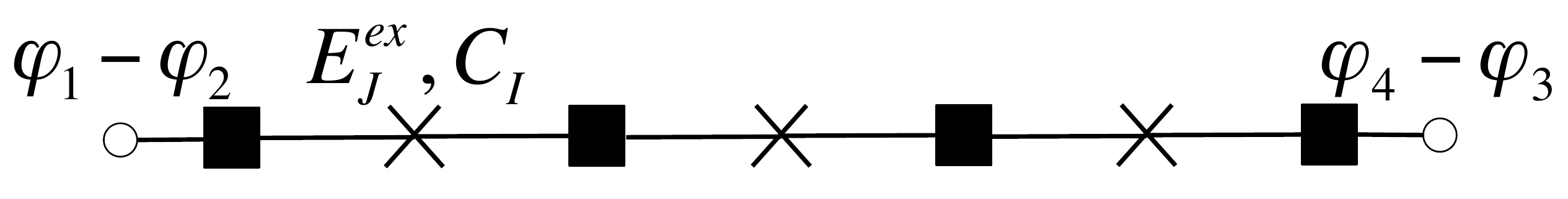}}}
\end{center}
\vspace*{-0.5cm} \caption{$\mathcal{H}_{\rm{eff}}$ in (\ref{H_eff}) describes a single chain of junctions with excitons as the tunnelling objects.} \label{ExcitonJJchain}
\end{minipage}
\end{figure}

Using $C_1/C_I <<1$, (\ref{H_KCMQ}) takes the approximate form 

\begin{eqnarray}
\mathcal{H} &\approx& \mathcal{O}(E_1) \sum_x n_{+}(x)^2 + E_I \sum_x n_{-}(x)^2 + \mathcal{O}(E_1) \sum_{x;y>0} n_{+}(x) n_{+}(x+y) \nonumber \\
&\,& -2 E_J \sum_x \cos \left[ \varphi_{+}(x+1) - \varphi_{+}(x) \right] \cos \left[ \varphi_{-}(x+1) - \varphi_{-}(x) \right],
\label{H_KCMQapprox}   
\end{eqnarray}
where $\varphi_\pm(x) = \left[\varphi_1(x) \pm \varphi_2(x) \right]/2$ and $n_\pm(x)\equiv n_1(x) \pm n_2(x)$, and terms $\mathcal{O}(C_1/C_I)E_I$ or smaller have been dropped. Employing degenerate perturbation theory, it is possible to find an effective Hamiltonian which describes the low energy dynamics of the system. Treating the Josephson term in (\ref{H_KCMQapprox}) as a perturbation and projecting into the subspace $n_{+}(x)=0, \, n_{-}(x)=0,\pm2$, one arrives at an effective Hamiltonian, $\mathcal{H}_{\rm{eff}}$, which describes the system as a single chain of junctions with excitons as the tunnelling objects instead of Cooper-pairs (see Fig.~\ref{ExcitonJJchain}):

\begin{equation}
\mathcal{H}_{\rm{eff}} =  4 E_I \sum_x n'_{-}(x)^2 - E_J^{\rm{ex}} \sum_x \cos \left[ \varphi'_{-}(x+1) - \varphi'_{-}(x) \right]. 
\label{H_eff}
\end{equation}
$\varphi'_{-}(x) = 2\varphi_{-}(x)$ and $n'_{-}(x) = n_{-}(x)/2$ are the phase of the macroscopic exciton wavefunction and the number of excitons, at position $x$, respectively, and are conjugate variables: $[\varphi'_{-}(x),n'_{-}(y)]=i\delta_{xy}$. The Josephson energy for the excitons is $E_J^{\rm{ex}} \equiv E_J^2/E_1$, while their charging energy is set by $E_I$. In the regime where $E_I$ is much less than $E_J^{\rm{ex}}$ (which requires $E_J >> E_I$, given that $E_1 >> E_J$), the excitons form a condensate. This is equivalent to saying that the excitonic junctions are in the `superfluid' regime, so that excitons can move freely along the chain. Assuming that this is so, the exciton condensate phase has small quantum fluctuations.

Now that the system is effectively described by a single chain of excitonic junctions, its low energy properties can be determined by appealing to results known from the study of conventional chains of Josephson junctions \cite{Pop}. It turns out that the ground state of such a chain can in fact be described by a single degree of freedom: the number of phase slips in the chain. Consider a chain of $N$ junctions with a phase drop of $\gamma$ across the chain, in a regime where the Josephson energy $E_J$ dominates the charging energy. If we neglect the charging energy for the moment, the classical ground state consists of $\gamma$ being equally distributed amongst all the junctions: $\theta_i = \gamma/N$, where $\theta_i$ is the phase difference across junction $i$. The assumption here is that $\gamma/N$ is small enough, so that  $\theta_i = \gamma/N$ is the minimum energy configuration. Given this, the potential of the chain is simply $E_0 = E_J\sum_{i=1}^{N} 1 - \cos \theta_i \approx E_J \gamma^2/2N$. However, phase slips can occur, where the phase of one junction, say the $j$th, changes as $\theta_j \rightarrow \theta_j + 2\pi$. Assuming that the phase bias is constant (i.e. $\sum_i \theta_i = \gamma$ must always be satisfied), the phases of the other junctions must change a little to accomodate the phase slip. As a result, the energy of the chain changes to $E_1 = E_J (\gamma - 2\pi)^2/2N$, and after $m$ phase slips, the energy is $E_m = E_J (\gamma - 2\pi m)^2/2N$. These energies correspond to parabolae centred at $\gamma = 2\pi m$, and the curves $E_m$ and $E_{m+1}$ are degenerate at $\gamma = (2m+1)\pi$. Restoring the charging energy - which introduces quantum phase fluctuations which can give rise to phase slips - lifts these degeneracies. A tight-binding model \cite{Matveev} gives rise to energy bands which are $2\pi$ periodic; the lowest band was confirmed in \cite{Pop}.     
   
Applying these ideas to Kitaev's device, we see that the lowest energy band can be written primarily as a function of the phase difference at the ends, $(\varphi_4 - \varphi_3) - (\varphi_1 - \varphi_2)$:

\begin{equation}
E=F\left( \varphi_4 - \varphi_3 + \varphi_2 - \varphi_1 \right) + \rm{error \, term},
\label{E_lowestband}
\end{equation} 
where $F$ is $2\pi$ periodic. The `error term' arises from uncorrelated currents, i.e. currents which are non-excitonic in nature, and is predicted to decrease exponentially as the number of junctions increases. If the first and third leads are connected so that $\varphi_1 = \varphi_3$, and similarly the second and fourth, the energy becomes $\pi$ periodic in $\varphi_2 - \varphi_1$, up to exponentially small corrections: $E \approx F\left( 2(\varphi_2 - \varphi_1) \right)$. $F\left( 2(\varphi_2 - \varphi_1) \right)$ has minima at $\varphi_2 - \varphi_1 = 0$ and $\varphi_2 - \varphi_1 = \pi$. As long as the barrier separating the minima is large enough, there are two ground states localized around these minima; these states can be identified with the logical states of a qubit (see Fig.~\ref{energy} in Sec.~\ref{VEQ}, where the same energy landscape arises for our `dual' construction). 

The `error term' leads to an exponentially small difference in the energy of the minima, which would inhibit dephasing. It is in this sense that the qubit is protected. However, a problem is that as the length of the device $N$ increases, the barrier height which is $\mathcal{O}(E_J^{ex}/N)$ decreases. This means that bit flips (tunnelling between the wells) becomes more of a headache, in the quest to suppress dephasing. As a consequence, the qubit would most likely have to run at very low temperatures, in order to minimize thermal fluctuations which would aid tunnelling. Also, the `mass' of the system, which is determined by $C_I$ (see the kinetic or charging energy term of $\mathcal{H}_{\rm{eff}}$ in (\ref{H_eff})) would have to be made sufficiently large, so that the system does not tunnel between the wells easily.

As was mentioned in the introduction, one may expect electric field fluctuations to pose a problem for Kitaev's current mirror qubit (KCMQ). Cooper-pair boxes suffer from charge fluctuations, and one may suspect that such fluctuations would affect the current mirror qubit as well. Electrons hopping back and forth between traps would effectively result in an external AC voltage being applied to the islands, which could result in uncorrelated currents flowing along the system, thus destroying the current mirror effect. Conventional wisdom says that, in the laboratory, electric noise is generally worse than magnetic noise: the relaxation and dephasing times for a flux qubit outperform a charge qubit, while phase qubits lag behind further still \cite{Clarke}. Thus, it seems reasonable to guess that a qubit analogous to KCMQ, based on magnetic degrees of freedom, would perform better in the laboratory. Furthermore, the magnetic analogue of Cooper-pair excitons - vortex-antivortex pairs, which are below referred to as `vortex excitons' - can arise naturally in a suitable 2D array of junctions. This provides another motivation for looking for a protected qubit which is `dual' to KCMQ.  

Kitaev also suggested measurement protocols and a universal set of one and two-qubit gates. These schemes involve connecting the qubit(s) to various circuits; details are contained in Sec.~\ref{M+G}, where the measurements and gates are adapted to the `dual' qubit that we describe next.

\section{A protected vortex exciton qubit: the `dual' of KCMQ} \label{VEQ}   

\begin{figure}[t]
\begin{minipage}{\columnwidth}
\begin{center}
\resizebox{.5\columnwidth}{!}{\rotatebox{0}{\includegraphics{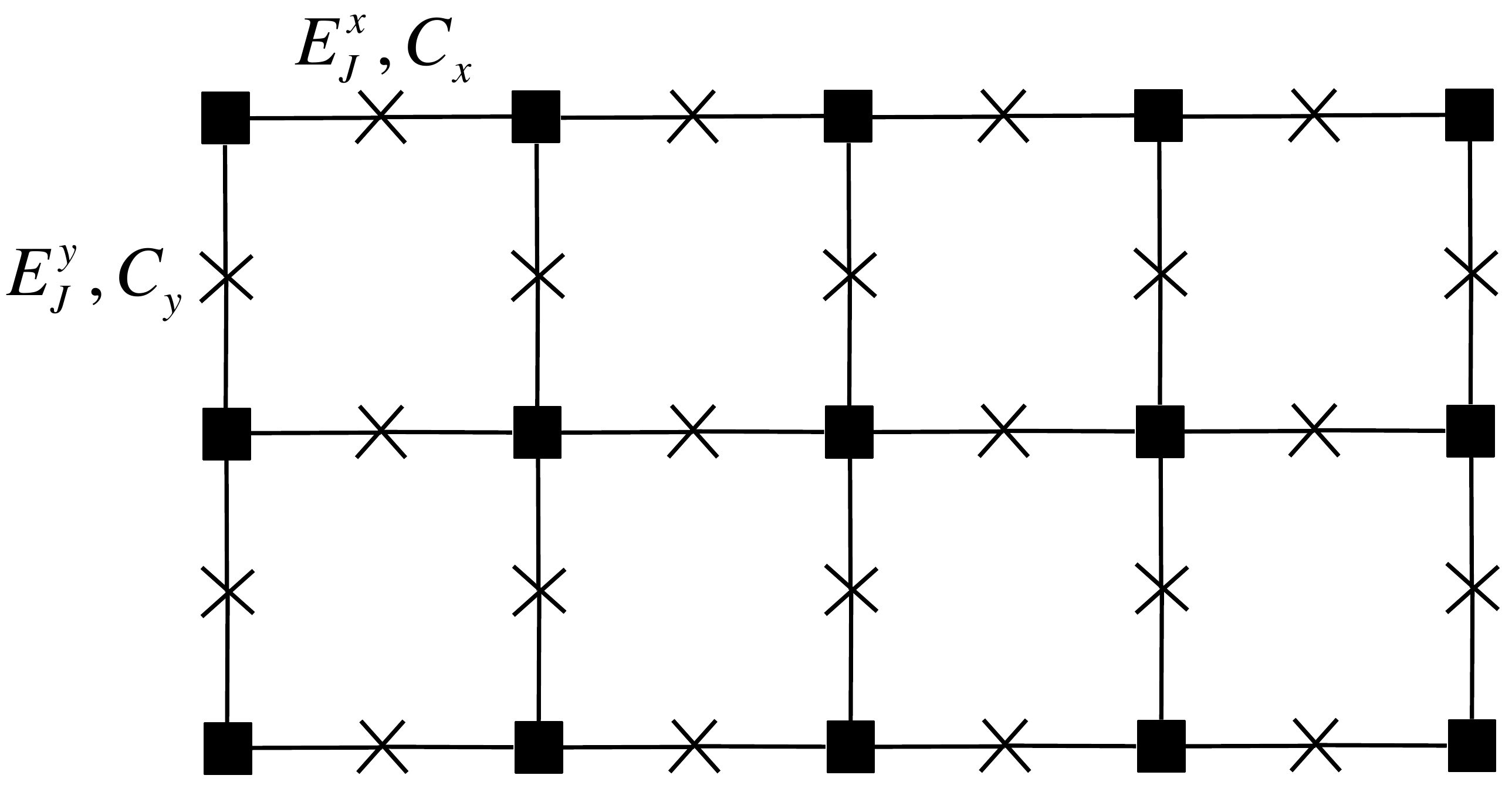}}}
\end{center}
\vspace*{-0.5cm} \caption{The Josephson junction array which gives rise to a protected vortex exciton qubit.} \label{VortexArray}
\end{minipage}
\end{figure}

The starting point for finding the `dual' of Kitaev's qubit is to employ a rectangular array of junctions, containing anisotropies in the junction capacitances as well as the Josephson couplings, as shown in Fig.~\ref{VortexArray}. Let us focus on the Josephson couplings first. We take the coupling in the $y$ direction to dominate that in the $x$ direction: $E_J^y >> E_J^x$. The reason for assuming this is that in a classical array (an array where the Josephson couplings dominate the corresponding charging energies) with isotropic couplings, the vortices in the array behave as a 2D Coulomb gas of charges, with the strength of their interaction set by the Josephson energy. With $E_J^y >> E_J^x$, it turns out that it costs more energy for a vortex-antivortex pair (`vortex exciton') to line up horizontally rather than vertically, in analogy with the allowed configurations of charge excitons in KCMQ.

It is convenient to start with a path integral representation of the partition function of the system \cite{F+S} \cite{Swanson} ($\hbar = 1$) 

\begin{equation}
\mathcal{Z} = \prod_i \int_0^{2\pi} d\phi_i \int_{\{\phi_i\}}^{\{\phi_i\}} D\{Q_j\}D\{\phi_j\} e^{-\mathcal{S}[\{Q_j(\tau)\},\{\phi_j(\tau)\}]},
\label{Z}
\end{equation}
where the Euclidean action is 

\begin{equation}
\mathcal{S}[\{Q_j(\tau)\},\{\phi_j(\tau)\}] \equiv \int_0^{\beta} d\tau \left( \frac{(2e)^2}{2}\sum_{i,j} Q_i(\tau) C_{ij}^{-1} Q_j(\tau) - i\sum_i Q_i(\tau) \dot \phi_i(\tau) - \sum_{\langle i,j \rangle} E_J^{ij} \cos \phi_{ij} (\tau) \right).  
\label{S} 
\end{equation}
In (\ref{Z}), $\int D\{\phi_j\}$ denotes integration over all paths that the phases $\{\phi_j\}$ can take satisfying $\phi_i(0) = \phi_i(\beta) = \phi_i$, and $\int D\{Q_j\} \equiv \prod_j \int DQ_j$ where    

\begin{equation}
\int DQ_j \equiv \underset{N \to \infty}{\lim} \frac{1}{(2\pi)^N} \sum_{Q_{j,0} = -\infty}^{\infty} \cdots \sum_{Q_{j,N-1} = -\infty}^{\infty},
\end{equation}
with $Q_{j,k}$ representing the dimensionless charge on island $j$ at discrete time $\tau_k = \epsilon k$ ($\epsilon$ is the time spacing). In the action (\ref{S}), $ \phi_{ij} \equiv  \phi_{i} -  \phi_{j}$, $\langle i,j \rangle$ denotes nearest neighbour sites $i$ and $j$ while $C_{ij}^{-1}$ denotes the inverse of the capacitance matrix. In the Villain approximation \cite{Villain} \cite{JKKN}, the Josephson energy contribution to $\mathcal{Z}$ can be written as

\begin{equation}
\sum_{\{ \vec{m}_{i,\tau} \} } \exp \left[ -\frac{\epsilon}{2} \sum_{i,\tau} E_J^x (\phi_{i+ \hat x,\tau} - \phi_{i,\tau} - 2\pi m_{i,\tau}^x)^2 + E_J^y (\phi_{i+ \hat y,\tau} - \phi_{i,\tau} - 2\pi m_{i,\tau}^y)^2 \right],
\label{JosEnVillForm} 
\end{equation}
where $\vec{m}_{i,\tau} = (m_{i,\tau}^x , m_{i,\tau}^y )$ is an integer-valued vector field associated with the links emerging from site $i$ in the $+\hat x$ and $+\hat y$ directions, with $\hat x , \hat y$ being the two lattice unit vectors. The crucial point is that the phases are now quadratic, which means that they can be integrated out and replaced by integer-valued charges residing on the dual lattice - the vortices. 

After some detailed calculations \cite{F+S}, the partition function becomes

\begin{equation}
\mathcal{Z} = \sum_{ \{ Q_{i,\tau} \} } \sum_{ \{ V_{i,\tau} \} } ' e^{- \mathcal{S}_{\rm{CCG}} \left( \{ Q_{i,\tau} \} , \{ V_{i,\tau} \} \right) } ,
\label{Z(Q,V)}
\end{equation}
with the coupled Coulomb gas action given by 

\begin{equation}
\mathcal{S}_{\rm{CCG}} \left( \{ Q_{i,\tau} \} , \{ V_{i,\tau} \} \right) = \epsilon \sum_{i,j,\tau} \frac{(2e)^2}{2} Q_{i,\tau} C_{ij}^{-1} Q_{j,\tau} + \pi  E_J^y V_{i,\tau} I'_{ij} V_{j,\tau} + i \frac{\sqrt{\beta}}{\epsilon} V_{i,\tau} \Theta_{ij} \partial_{\tau} Q_{j , \tau} + \frac{1}{4\pi \epsilon^2 E_J^x} \partial_{\tau} Q_{i , \tau} I_{ij} \partial_{\tau} Q_{j , \tau}.
\label{Sccg}  
\end{equation}
The variables $V_{i,\tau}$ are integer-valued and represent vortex degrees of freedom, which live on the dual lattice; the prime on the sum in (\ref{Z(Q,V)}) indicates that only neutral configurations of vortices enter $\mathcal{Z}$. The matrix $I_{ij}$ determines the kinetic contribution of the charges ($\partial_{\tau} Q_{i,\tau}\equiv Q_{i,\tau} - Q_{i,\tau - \epsilon}$) and is given by  

\begin{equation}
I_{ij}(\beta) = \int_0^{\pi} dq \frac{e^{- \vert x_i - x_j \vert s} \cos \left[(y_i - y_j) q \right]}{\sinh (s)} ,
\end{equation}
where $\sinh (s) = \sqrt{\left[1 + \beta (1 - \cos (q) ) \right]^2 - 1}$ and $\beta = E_J^y / E_J^x $ is the Josephson energy anisotropy. $I_{ij}$ also determines the interaction between the vortices, which is $I_{ij}' = I_{ij} - I_{ii}$. Charges and vortices interact via the potential $\Theta_{ij}$, which is an anisotropic version of the potential stated in \cite{F+S} for isotropic arrays.  

Given the emergence of vortex degrees of freedom in $\mathcal{S}_{\rm{CCG}}$, and the almost perfect symmetry that exists between charges and vortices in $\mathcal{S}_{\rm{CCG}}$, it is tempting to think that a \textit{dual} description of the system exists. Namely, a representation in terms of vortices $\{ V_i \}$ and phases $\{ \theta_i \}$ (residing on the dual lattice) conjugate to the vortices. Indeed, for arrays in which the Josephson energies dominate the corresponding charging energies (such as the array we consider - see below), vortices are the relevant dynamical degrees of freedom, and it is convenient to analyse a Hamiltonian defined in terms of vortices and their conjugate phases \cite{vanWees91} \cite{Choi94}. Such a Hamiltonian can be written for the array we consider:  

\begin{equation}
\mathcal{H}^v = \pi E_J^y \sum_{i,j} V_{i} I'_{ij} V_{j} - \frac{2}{\pi^2} E_C^y \sum_i \cos \left( \theta_{i+ \hat x} - \theta_{i} \right) - \frac{2}{\pi^2} E_C^x \sum_i \cos \left( \theta_{i+ \hat y} - \theta_{i} \right),
\label{Hdual}
\end{equation}
where the vorticities and phases satisfy the commutation relation $\left[ \theta_i,V_j \right] = i \delta_{ij}$. 

Just as the vortices emerged from the Villain approximation of the phases $\{ \phi_i \}$, it is clear that a Villain approximation of the dual phases $\{ \theta_i \}$ in (\ref{Hdual}) would yield the electric charging energy contribution to $\mathcal{S}_{\rm{CCG}}$. $C_{ij}$ - and in particular, the charging energies associated with the junction capacitances - fixes the dual Josephson couplings to be $E_{JD}^y \equiv \frac{2}{\pi^2} E_C^x = \frac{2}{\pi^2} \frac{e^2}{2C_x}$ and $E_{JD}^x \equiv \frac{2}{\pi^2} E_C^y = \frac{2}{\pi^2} \frac{e^2}{2C_y}$. Of course, (\ref{Hdual}) cannot account for the last term in (\ref{Sccg}), which breaks the perfect symmetry between the charges and the vortices, and reflects the spin-wave excitations of $\{ \phi_i \}$. Such spin-waves can in general be a source of dissipation for vortices. However, we assume that the self-capacitances of the islands are neglegible compared to the junction capacitances, since experiments bear this out \cite{F+vdZant}. For such arrays, the spin-wave dispersion only has an optical branch \cite{F+vdZant}, which presumably makes the spin-wave-vortex coupling irrelevant for vortex dynamics at low temperatures.    
         
We would like to `mimic' the physics of the Kitaev qubit, where Cooper-pair excitons are in a superfluid regime and move freely along an array. The analog of these charge excitons here are vortex excitons, which can bind naturally in a classical array once the temperature drops below the Berezinskii-Kosterlitz-Thouless (BKT) temperature, $T_{\rm{BKT}}$ \cite{Berez,KT}. The BKT transition is associated with the binding and unbinding of vortex-antivortex pairs; for $T<T_{\rm{BKT}}$, it becomes free-energetically more favourable for a vortex and antivortex to pair up, rather than to remain as free entities. 

\begin{figure}[t]
\begin{minipage}{\columnwidth}
\begin{center}
\resizebox{.5\columnwidth}{!}{\rotatebox{0}{\includegraphics{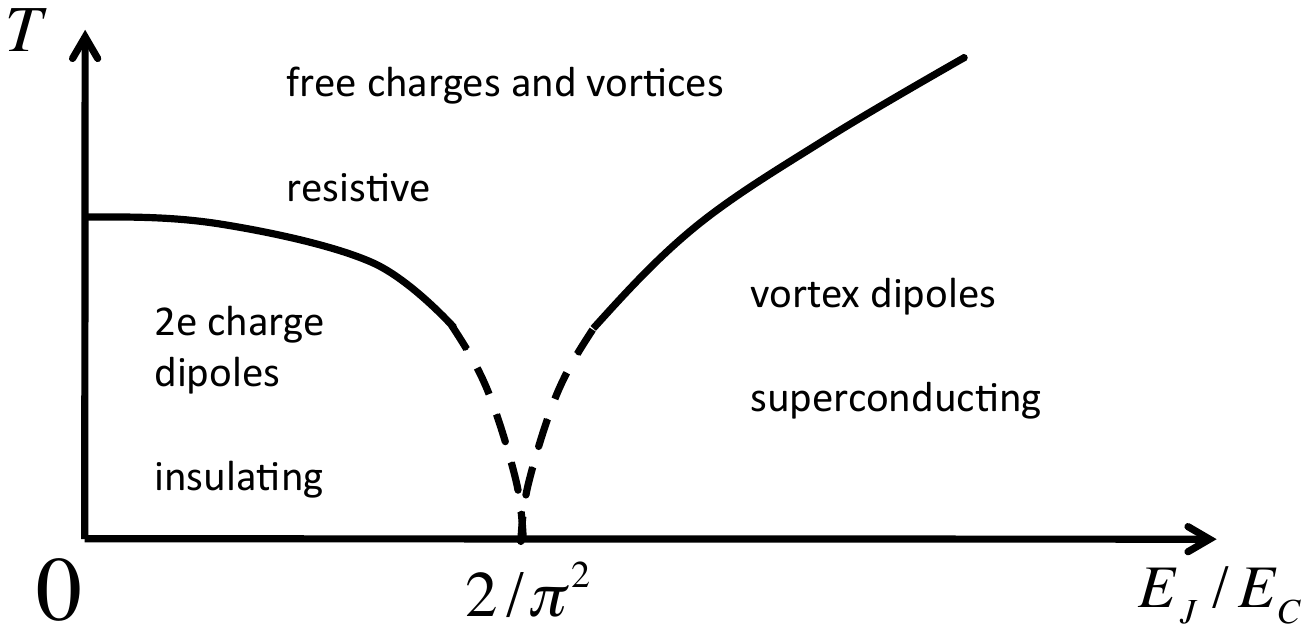}}}
\end{center}
\vspace*{-0.5cm} \caption{The phase diagram for an isotropic array (see \cite{F+S}).} \label{PhaseDiagram}
\end{minipage}
\end{figure}

In an array in which quantum effects are relevant, i.e. when the charging energies are not negligible, the phase diagram for an isotropic array will be as shown in Fig.~\ref{PhaseDiagram}. In such cases, the vortex-unbinding $T_{\rm{BKT}}$ depends on the ratio of the Josephson and charging energies. For us, the presence of anisotropies,  both in Josephson and charging energies, means that the phase diagram acquires `extra dimensions' (for example, one `extra dimension' will be the Josephson energy anisotropy parameter, $\beta$). We will assume that whatever our required array parameters, we can place the system in an area of the phase diagram where vortices and antivortices are bound together. Below, we will outline additional arrays of junctions for measurements and gates, in which we will require the existence of free vortices. As can be seen in Fig.~\ref{PhaseDiagram}, this means that it is undesirable to run the qubit in an ultra-low temperature environment, since the only two phases available to the system at $T \approx 0$ are the vortex dipole and charge dipole phases. 

\begin{figure}[t]
\begin{minipage}{\columnwidth}
\begin{center}
\resizebox{.5\columnwidth}{!}{\rotatebox{0}{\includegraphics{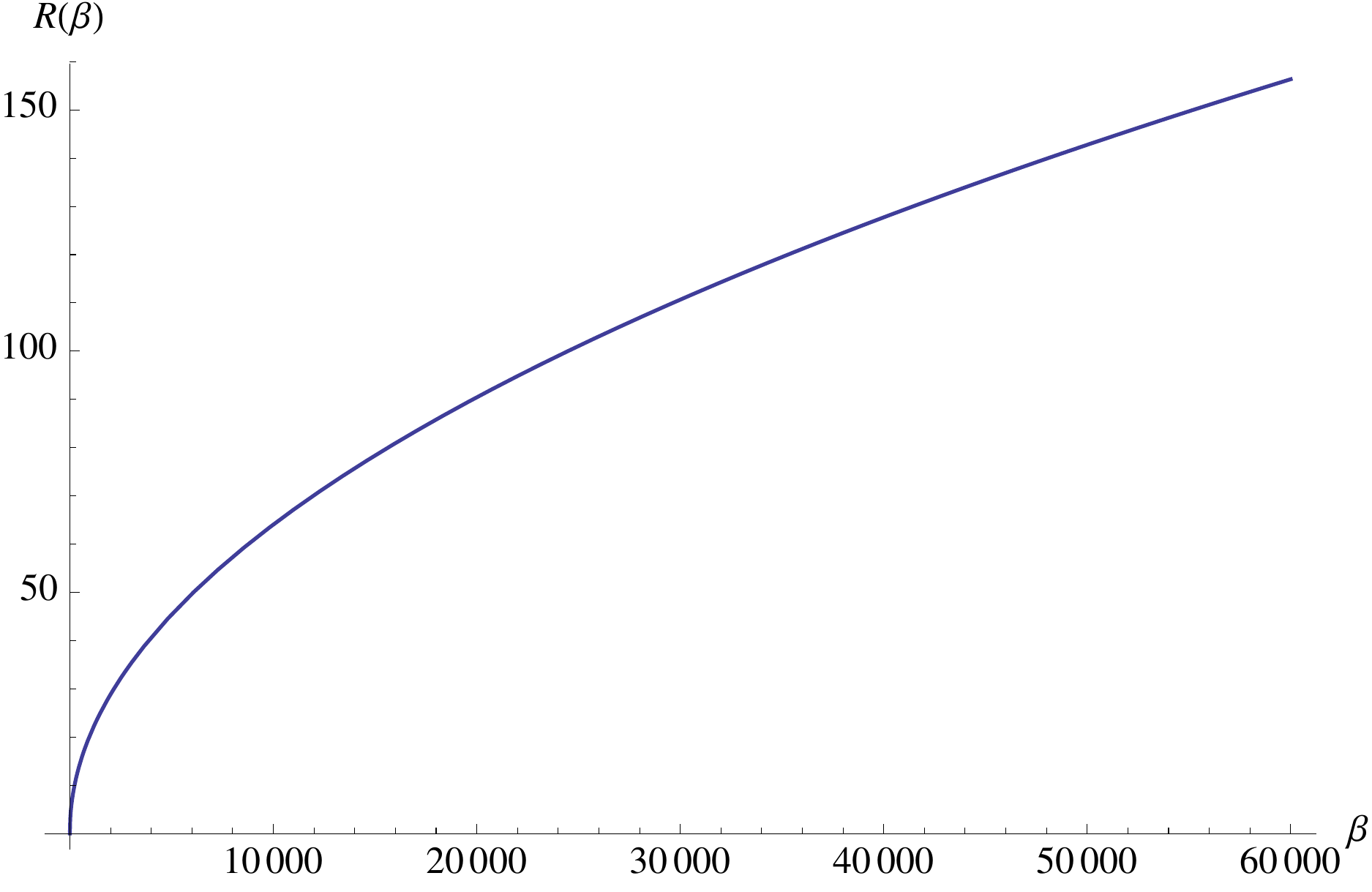}}}
\end{center}
\vspace*{-0.5cm} \caption{Ratio of energies required for a nearest neighbour vortex exciton to form in the $x$ direction compared to $y$ direction, $R(\beta)$, as a function of Josephson coupling anisotropy $\beta = E_J^y / E_J^x$.} \label{Ratio}
\end{minipage}
\end{figure}

In KCMQ, the charge excitons are `lined up' in the vertical ($y$) direction, and we have ensured that the vortex excitons also orient themselves in this manner, because we said at the outset that $E_J^y >> E_J^x$. The ratio of the energy required to form nearest neighbour vortex dipoles in the $x$ direction ($+ -$ or $- +$) compared to in the $y$ direction ($\pm$ or $\mp$) is given by $I'(1,0;\beta) / I'(0,1;\beta) \equiv R(\beta)$ \footnote{$I_{ij}(\beta) \equiv I(x_i - x_j , y_i - y_j ; \beta)$, where distances are measured in units of the lattice spacing.}, and this is plotted in Fig.~\ref{Ratio}. It is clear that for large enough anisotropy (i.e. $ \beta = E_J^y / E_J^x >> 1$), it is energetically much more favourable for vortex excitons to be aligned vertically rather than horizontally.        

In KCMQ, the motion of charges is restricted to one dimension, along the length of the array ($x$ direction). In order to ensure that vortices can also only move along the length of the array that we are considering, we require the Josephson tunnelling of the vortices in the $y$ direction to be suppressed. This can be achieved by making the inertial resistance of the vortices in the $y$ direction to be much larger than in the $x$ direction i.e. by enforcing $C_x >> C_y \rightarrow E_C^y >> E_C^x$. Furthermore, since we want the array to be in the regime of phase coherence where vortices are the relevant degrees of freedom, we require that the Josephson energies dominate the corresponding charging energies, i.e. $E_J^i >> E_C^i, i\in \{ x,y \}$. In particular, the condition $E_J^y >> E_C^y$ ensures that uncorrelated hopping of the vortices is too costly, although correlated motion of vortex excitons is possible, as we shall see below.

Neglecting the vortex tunnelling in the $y$ direction, we can write the Hamiltonian (\ref{Hdual}) as

\begin{equation}
\mathcal{H}^v \approx \pi E_J^y \sum_{i,j}  V_{i} I'_{ij} V_{j} - \frac{2}{\pi^2} E_C^y \sum_i \cos \left( \theta_{i+ \hat x} - \theta_{i} \right),
\end{equation} 
and using the notation in \cite{Choi}, this can be rewritten as

\begin{equation}
\mathcal{H}^v = \pi E_J^y \sum_{l,l',x,x'} V_{l}(x) I'_{ll'}(x,x') V_{l'}(x') - \frac{2}{\pi^2} E_C^y \sum_{l,x} \cos \left( \theta_{l} (x+1) - \theta_{l} (x) \right) ,
\end{equation} 
where $l=1,2$ denotes the $y$ coordinate of the dual lattice sites, while $x$ denotes the coordinate along the length of the array. Since there is now a one-to-one correspondence between the essential features of KCMQ and our dual construction, $\mathcal{H}^v$ is well approximated by (compare with (\ref{H_KCMQapprox}))

\begin{eqnarray}
\mathcal{H}^v &\approx& \mathcal{O}[-I'(1,0; \beta )] E_J^y \sum_x V_{+}(x)^2 + \mathcal{O}[-I'(0,1; \beta )] E_J^y \sum_x V_{-}(x)^2 + \mathcal{O}[-I'(1,0; \beta )] E_J^y \sum_{x;y>0} V_{+}(x) V_{+}(x+y) \nonumber \\
&\,& -\frac{4}{\pi^2} E_C^y \sum_x \cos \left[ \theta_{+}(x+1) - \theta_{+}(x) \right] \cos \left[ \theta_{-}(x+1) - \theta_{-}(x) \right],
\label{Hdualapprox}   
\end{eqnarray}
where $V_\pm (x) = V_1 (x) \pm V_2 (x)$ and $\theta_\pm (x) = [\theta_1 (x) \pm \theta_2 (x)]/2$.   

\begin{figure}[t]
\begin{minipage}{\columnwidth}
\begin{center}
\resizebox{.37\columnwidth}{!}{\rotatebox{0}{\includegraphics{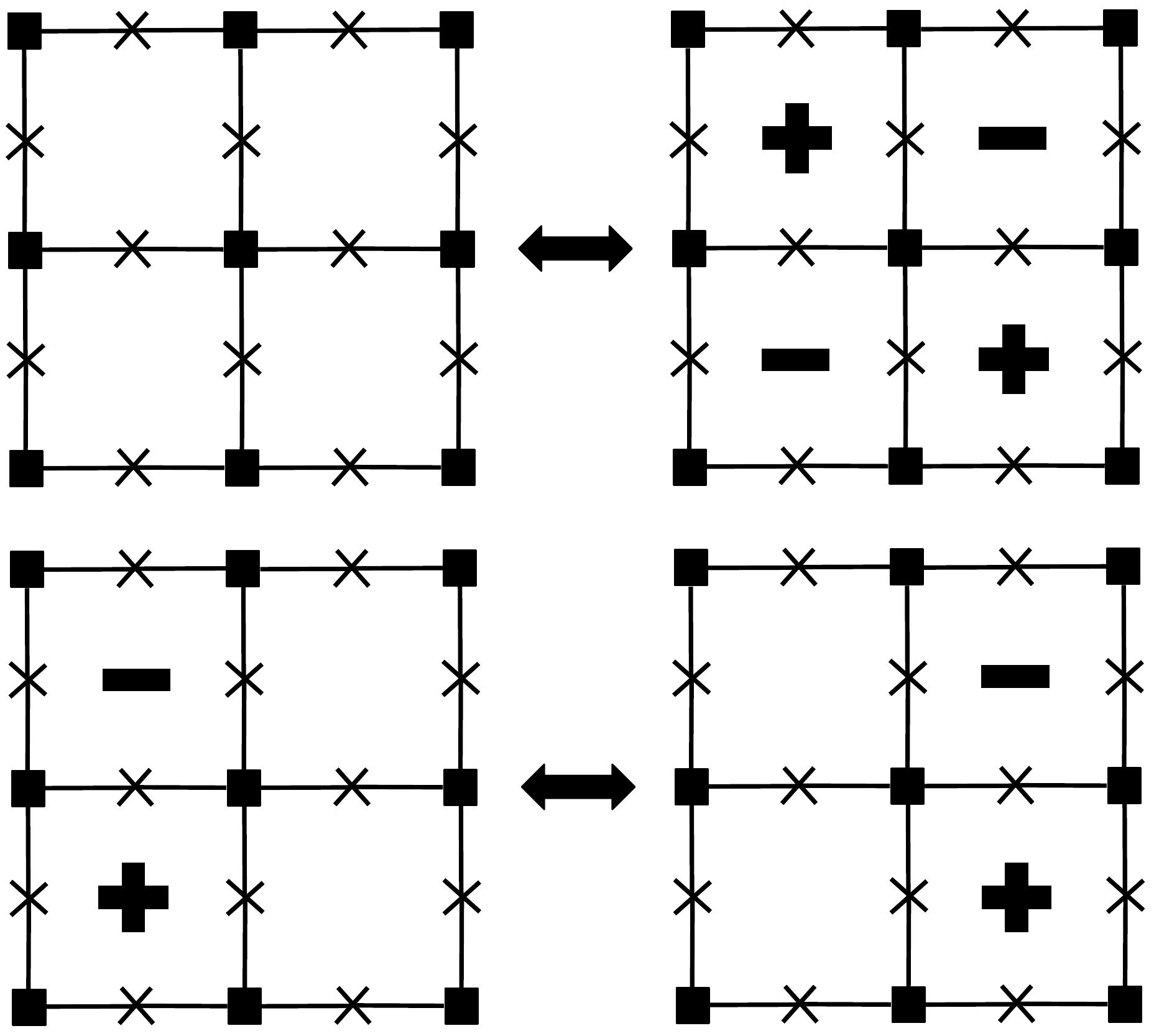}}}
\end{center}
\vspace*{-0.5cm} \caption{Examples of vortex exciton tunnelling.} \label{Tunn}
\end{minipage}
\end{figure}

We are interested in the low energy character of the system, and since the energy scale for horizontal excitons ($ \mathcal{O}[-I'(1,0; \beta )] E_J^y$) dominates the other energy scales, we can project (\ref{Hdualapprox}) into the subspace $V_+(x) =0, \, V_-(x) =0,\pm2$. Doing so yields an effective Hamiltonian 

\begin{equation}
\mathcal{H}_{\rm{eff}}^v \approx  \mathcal{O}\left[ -4I'(0,1; \beta ) \right] E_J^y  \sum_x V_{-}^{'} (x)^2 - E_C^{\rm{ex}} \sum_{x} \cos \left[ \theta_{-}^{'} (x+1) - \theta_{-}^{'} (x) \right],
\label{EffHam}  
\end{equation}
where $\theta_{-}^{'} (x) = 2 \theta_{-} (x)$, $V_{-}^{'} (x) = V_{-} (x)/2$ and $E_C^{\rm{ex}} \equiv \frac{4}{\pi^4} \frac{ {E_C^y }^2 }{ \mathcal{O} \left[ -I'(1,0; \beta ) \right] E_J^y }$. $\mathcal{H}_{\rm{eff}}^v$ corresponds to a 1D chain of Josephson junctions with vertically aligned vortex excitons as the tunnelling objects. In other words, the effective low energy degrees of freedom are vortex-antivortex pairs at position $x$, characterized by quantum number $V_{-}^{'} (x)$, and these excitons tunnel under the influence of a Josephson coupling energy $E_C^{\rm{ex}}$, as shown schematically in Fig.~\ref{Tunn}. This is precisely the scenario in KCMQ, with vortices replaced by Cooper-pairs. For the system to be in the `superfluid' phase where the fluctuations in the phases $\theta_{-}^{'}$ are small and the dipoles tunnel freely, the tunnelling strength must dominate the vortex `charging energy', i.e. $E_C^{\rm{ex}} >> \mathcal{O}\left[ -4 I'(0,1; \beta ) \right] E_J^y $, which leads to the requirement 

\begin{equation}
\left( \frac{E_C^y}{E_J^y} \right)^2 >> \pi^4 \mathcal{O}\left[ I'(0,1; \beta ) \right] \mathcal{O}\left[ I'(1,0; \beta ) \right].
\end{equation}
There exists parameter regimes where the condition immediately above and the condition $\mathcal{O}\left[ -I'(1,0; \beta ) \right] E_J^y  >> (4/\pi^2) E_C^y$ (so that the projection into the subspace with $V_+(x) =0 \, , V_-(x) =0,\pm2$ is valid) can be satisfied.

\begin{figure}[t]
\begin{minipage}{\columnwidth}
\begin{center}
\resizebox{.5\columnwidth}{!}{\rotatebox{0}{\includegraphics{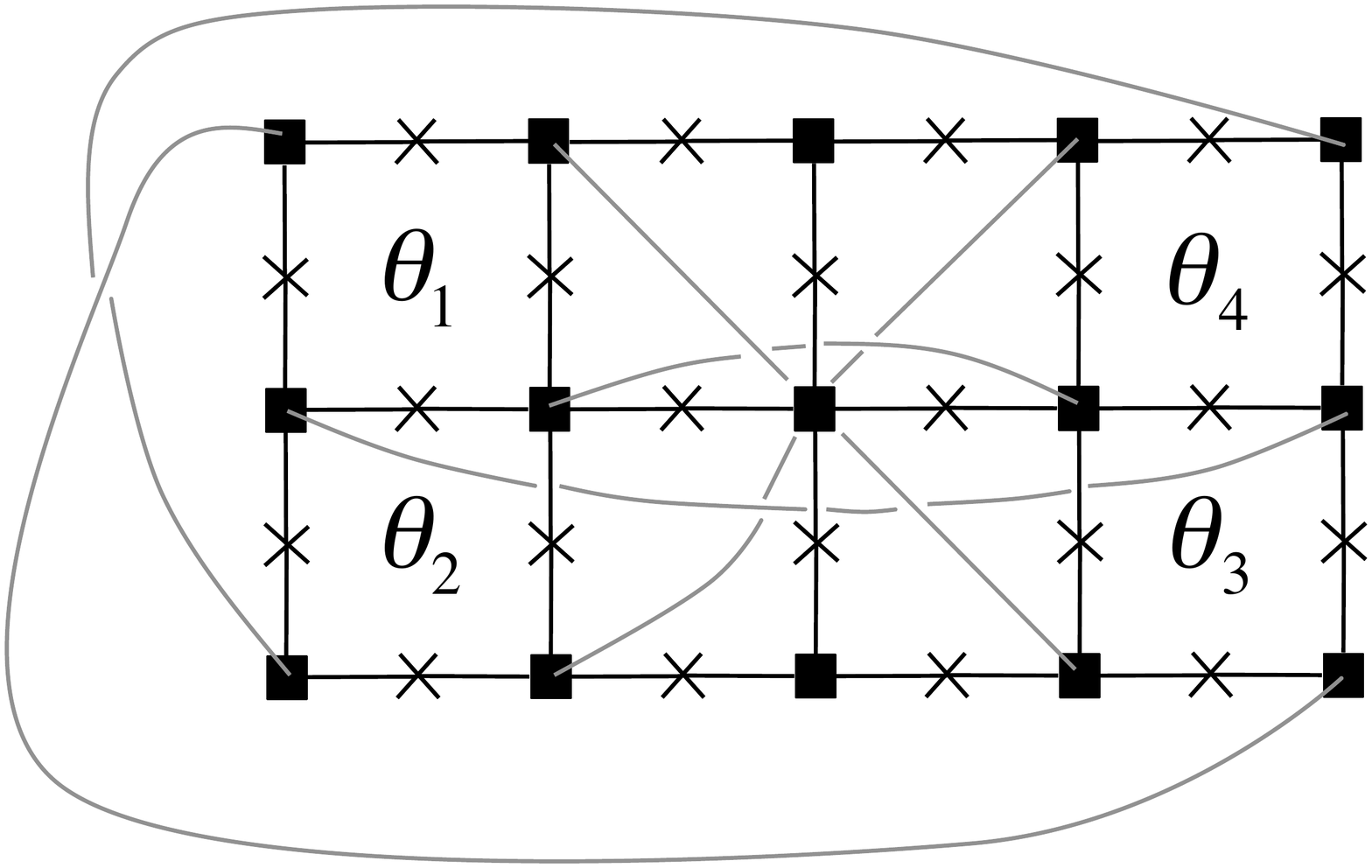}}}
\end{center}
\vspace*{-0.5cm} \caption{The superconducting wires that need to be attached to establish the boundary conditions needed for a $\pi$-periodic energy landscape.} \label{bcs}
\end{minipage}
\end{figure}

The energy of the system is described by an equation analogous to (\ref{E_lowestband}), with the replacement $\varphi_i \rightarrow \theta_i$, where $\theta_i$ are the corner phases of the array (see Fig.~\ref{bcs}). The existence of an `error term' is completely analogous to the `error term' identified by Kitaev. In addition to the tunnelling of vortex excitons, rare uncorrelated tunnelling of individual vortices may also occur. For such a process to provide a net contribution to vortex current (and hence a contribution to the energy, since the derivative of energy with respect to suitable phases yields a current), a vortex must make its way from one end of the array to the other. If the length of the array is $N$ and the probability of a vortex to hop to an adjacent site is $p$ - where $p$ is small - then the probability of a vortex to make its way from one end to the other is $p^N$. Hence, the `error term' decreases exponentially in the length of the array. 

\begin{figure}[t]
\begin{minipage}{\columnwidth}
\begin{center}
\resizebox{.5\columnwidth}{!}{\rotatebox{0}{\includegraphics{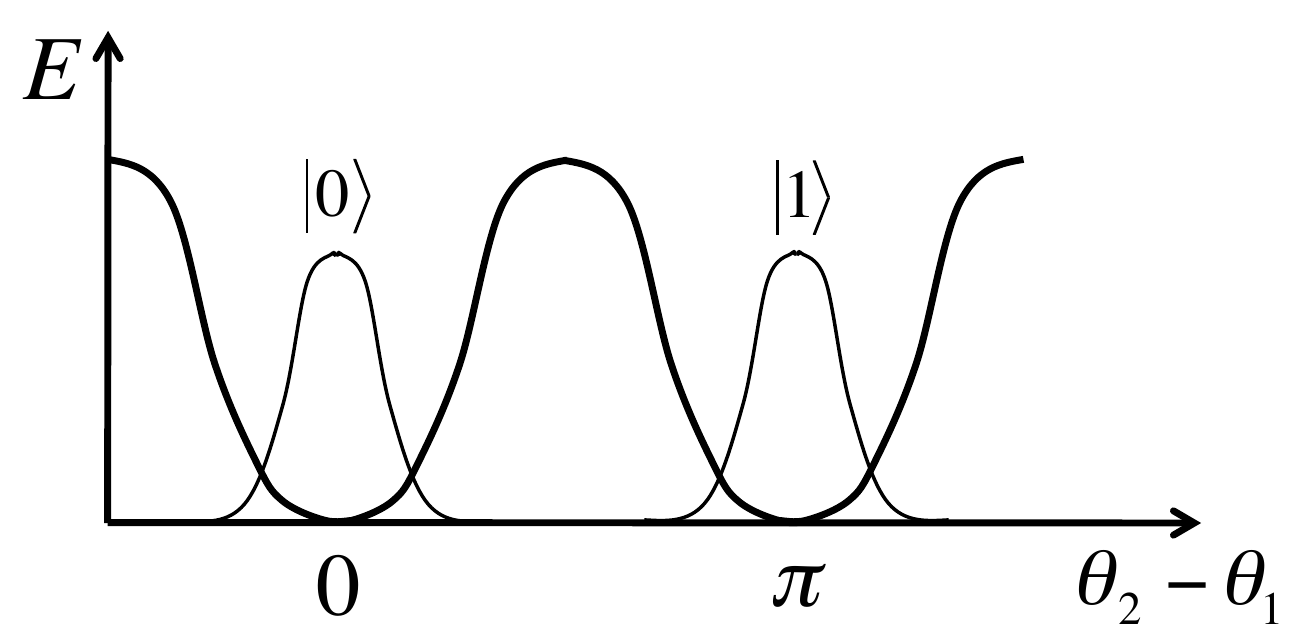}}}
\end{center}
\vspace*{-0.5cm} \caption{The $\pi$-periodic energy and the qubit basis states.} \label{energy}
\end{minipage}
\end{figure}

Kitaev arrived at a $\pi$-periodic energy by connecting the leads of his device diagonally. To establish a $\pi$-periodicity here, the phases of the top left plaquette and the bottom right plaquette must be identified, and similarly, the bottom left and the top right phases must be set equal too. Attaching superconducting wires between the islands in the manner depicted in Fig.~\ref{bcs} will do the job. The wires equalize the phases of the superconducting islands that they connect. Hence, if one were to track the winding of the superconducting phases of the top left and bottom right plaquettes, one would find that the windings (or vorticities) would be the same, which implies $\theta_1 = \theta_3$ (and similarly $\theta_2 = \theta_4$). The $\pi$-periodic energy that results from the boundary conditions, and the states localized around the two minima to be used as the qubit basis, are shown in Fig.~\ref{energy}. 

\section{Measurements and gates} \label{M+G}

In this section, we adapt the measurement schemes and the universal set of gates for KCMQ to our construction. The measurements and gates proposed by Kitaev are: 

\begin{enumerate}
\item Measurement in the computational basis
\item Measurement in the dual basis
\item The one-qubit unitary $ R(\pi/4) = \exp \left( i(\pi/4)Z \right) $ and its inverse
\item The two-qubit unitary $ R_2(\pi/4) = \exp \left( i(\pi/4) Z_1 Z_2 \right) $ and its inverse
\item The one-qubit unitary $ R(\pi/8) = \exp \left( i(\pi/8)Z \right) $ and its inverse
\end{enumerate}

\begin{figure}[t]
\begin{minipage}{\columnwidth}
\begin{center}
\resizebox{.5\columnwidth}{!}{\rotatebox{0}{\includegraphics{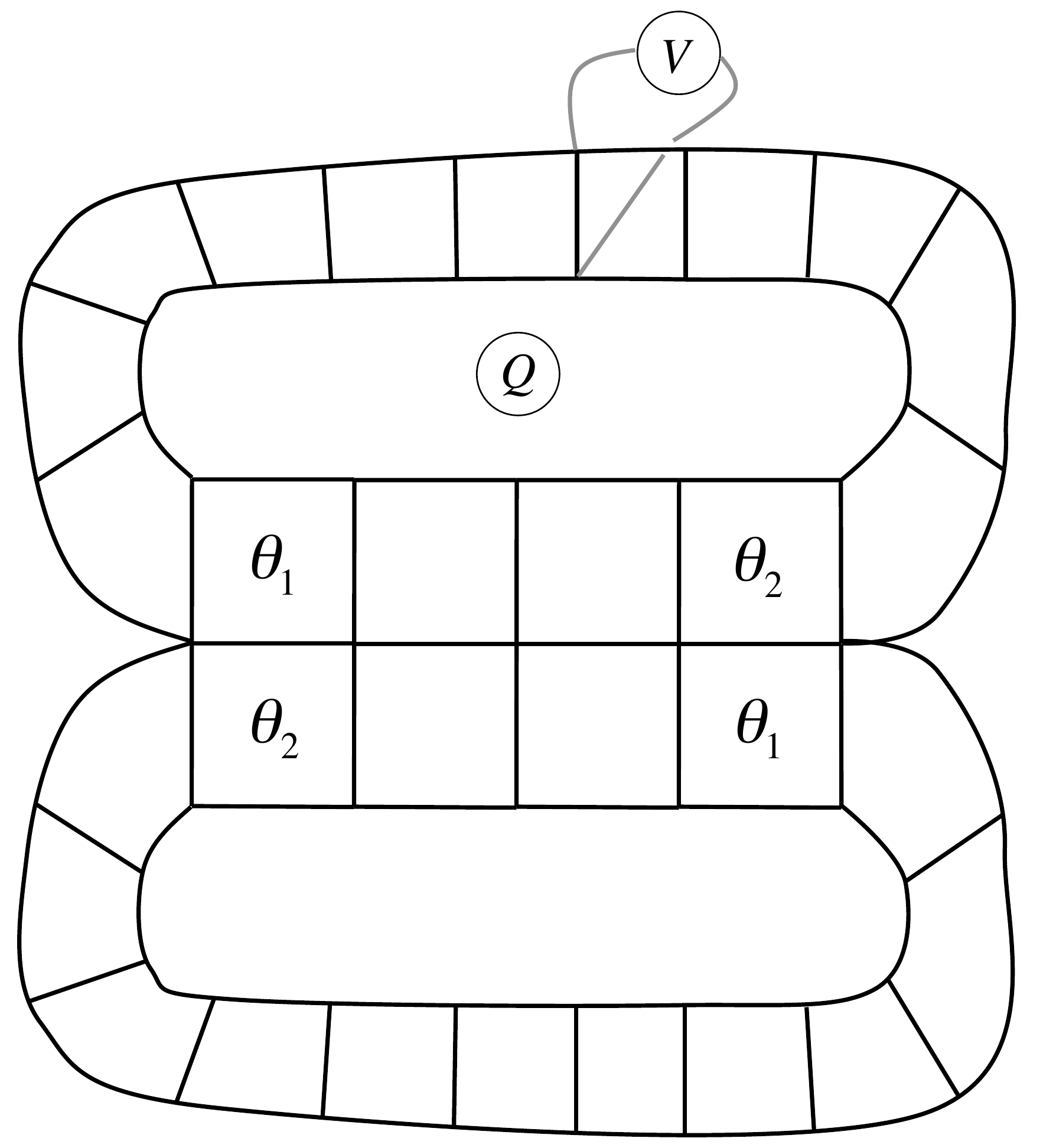}}}
\end{center}
\vspace*{-0.5cm} \caption{Scheme for measurement in the computational basis.} \label{CompMeas}
\end{minipage}
\end{figure}

A method of measuring in the computational basis $\{ \vert 0 \rangle ,   \vert 1 \rangle \}$ is sketched in Fig.~\ref{CompMeas}. Two `tracks' of junctions are attached onto the qubit, with parameters in the tracks chosen such that free or individual vortices can exist within them. Assuming the Josephson energies dominate the charging energies, the phase of the system in the `tracks' will depend on the ratio of the Josephson energies to the temperature. Since the small but finite temperature will be fixed, by tweaking the Josephson energies, the existence of free vortices can be guaranteed. The vortex current in the top loop depends on the external electric charge $Q$ threading the loop and the phase drop across the qubit, which is either $0$ or $\pi$. This is a manifestation of the Aharonov-Casher effect \cite{AC} which has already been seen in arrays \cite{Elion}. Thus, a measurement of the vortex current - which is equivalent to a voltage measurement, as shown in Fig.~\ref{CompMeas} - would tell us the state of the qubit.       

A dual basis ($\vert \pm \rangle = (\vert 0 \rangle \pm \vert 1 \rangle)/\sqrt{2} $) measurement is sketched as follows. In analogy with the dual measurement in KCMQ, which is explained in Appendix \ref{DualBasisMeas}, we require a parity measurement, which requires a `vortex charge qubit'. Similar to a conventional charge qubit, two equal and oppositely charged vortices sitting next to each other will store some `vortex charging energy'. The device we have in mind is similar to a dual DC SQUID which is discussed in Appendix \ref{VortexDCSQUID}. For concreteness, say the vortex on the left plaquette has charge $-N$ while its neighbour has charge $+N$. The precise form of the charging energy is given by the vortex-vortex interaction, which is discussed in Sec.~\ref{VEQ}. For short distances, the interaction is not easy to work with. Nonetheless, the fact that the charging energy is symmetric under $N \rightarrow -N$ and must tend to zero as $N \rightarrow 0$ are the only pieces of information that we need for our purposes. 

Since the energy is symmetric and is not simply constant, this device has properties reminiscent of a conventional charge qubit. If a vortex charge bias of $1/2$ (in units of the flux quantum) is applied to one of the plaquettes and $-1/2$ is applied to the other plaquette, the states $N=0$ and $N=1$ will become degenerate. However, vortex tunnelling (which can be controlled by electric charge $Q$, see Fig.~\ref{DualSquid}) will break the degeneracy, resulting in a spectrum that should be qualitatively similar to that of a charge qubit, at least for the lowest band. Hence, we call this device a `vortex charge qubit'.

A parity measurement now follows in complete analogy to the scheme for KCMQ. The first step is to cut the wires that equalize the phases $\theta_1$ and $\theta_3$. This results in a phase difference of $\theta_3 - \theta_1 =  2( \theta_2 - \theta_1 ) \equiv 2 \Theta$ between the free ends, which is conjugate to the vortex charge operator $n/2$. We now take a vortex charge qubit and apply the free ends of the system we are considering across it. Since the charge qubit has a unit period energy (with respect to bias charge), so does the ground state ‘vortex voltage’ (an electrical current). This means that if we apply a vortex charge bias of $\sim 1/4$, then measurement of the vortex charge qubit voltage will yield the parity of $n$.     

An unprotected rotation $R(\pi/8)$ can be realized using the setup in Fig.~\ref{CompMeas} with a few modifications: external charge and voltage measurement are not needed, while a switch of some sort should be inserted in the top track. This switch could be a dual or vortex DC SQUID, a realization of which is sketched in Appendix \ref{VortexDCSQUID}. When the switch is turned on, the qubit becomes coupled to the vortex DC SQUID with coupling term $-E_C \cos (\theta_2 - \theta_1) = -E_C Z$. The switch should thus be turned on for a certain amount of time to implement the gate. We assume that when the switch is inserted into the top track, parameters are chosen such that dual phase locking occurs, so that the coupling term $-E_C Z$ is the only addition to the Hamiltonian.   

\begin{figure}[t]
\begin{minipage}{\columnwidth}
\begin{center}
\resizebox{.7\columnwidth}{!}{\rotatebox{0}{\includegraphics{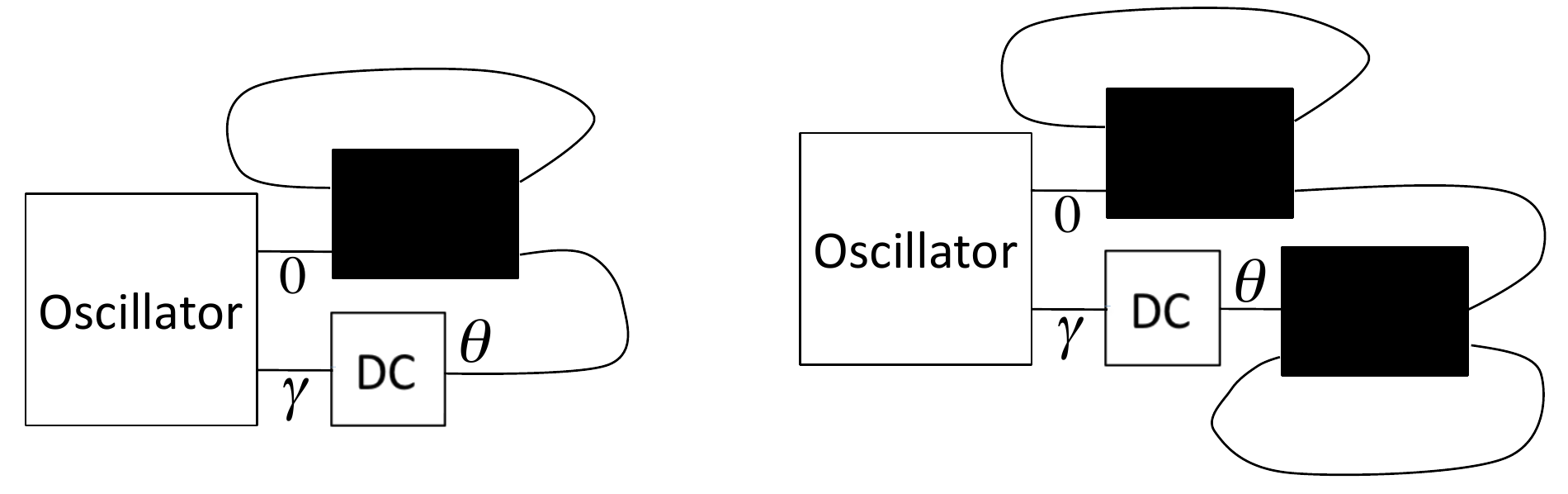}}}
\end{center}
\vspace*{-0.5cm} \caption{Left: Circuit for implementing $R(\pi/4)$. Right: Circuit for implementing $R_2(\pi/4)$. $\gamma$ is the phase of the oscillator and $\theta$ is the phase drop across the qubit(s).} \label{ProtectedGates}
\end{minipage}
\end{figure}

A fault-tolerant realization of the gate $R(\pi/4)$ was sketched by Kitaev. The idea is to use a continuous-variable error-correcting code \cite{GKP}, in which switching on a coupling between a qubit and a harmonic oscillator results in the qubit becoming encoded into the Hilbert space of the oscillator. The oscillator then picks up a qubit-state dependent phase factor, which is resilient to experimental imperfections such as timing errors or variations in coupling strength, due to the properties of the code \cite{Brooks}. When the coupling is switched off, the qubit and oscillator become disentangled, with $R(\pi/4)$ having been applied to the qubit. Similarly, $R_2(\pi/4)$ can be applied to two $0$-$\pi$ qubits by connecting them in series to an oscillator. 

In order to use the fault-tolerant gate $R(\pi/4)$ for the qubit we consider, we require a vortex $LC$ oscillator, in contrast to a conventional electric $LC$ oscillator which is needed for KCMQ. Furthermore, for gate errors to be exponentially suppressed, the impedance of such an oscillator should be large \cite{Brooks}. The construction of an oscillator with this property is sketched in Appendix~\ref{VortexOsc}. A vortex DC SQUID can operate as a switch, which is needed to mediate the coupling between the qubit and oscillator. The circuits corresponding to the implementation of $R(\pi/4)$ and $R_2(\pi/4)$ are shown in Fig.~\ref{ProtectedGates}.

\section{Conclusions} \label{Conclusions}
In this work, we have presented a qubit which is `dual' to a $0$-$\pi$ current mirror qubit (KCMQ) proposed by Kitaev \cite{KitaevJJ}. We find that in a suitable array of Josephson junctions, the physics of KCMQ can arise, with vortex and dual phase degrees of freedom replacing the Cooper-pair and superconducting phase degrees of freedom in KCMQ. As in KCMQ, a judicious choice of dual phase boundary conditions leads to the emergence of two nearly degenerate ground states, which can be used as a qubit. The splitting of the ground states is exponentially small in the length of the array, and we expect this to remain the case even in the presence of moderate disorder in array paramaters, as has been demonstrated for KCMQ \cite{Dempster}. Such a qubit thus has an in-built fault-tolerance which suppresses dephasing. We have also adapted the scheme for universal fault-tolerant quantum computation proposed by Kitaev to the dual qubit that we consider. This has required the construction of novel circuit elements, such as a vortex harmonic oscillator.   

One motivation for constructing such a dual qubit is that electric field noise may impair the current mirror effect which KCMQ relies on. On the other hand, magnetic field noise is less of an issue in the laboratory, and so a `dual' of KCMQ may well perform more favourably than KCMQ in practice. Another motivation is that the magnetic analogue of Cooper-pair excitons (which are present in KCMQ), namely vortex anti-vortex pairs, can arise naturally in low temperature arrays where Josephson energies are dominant. 

\section{Acknowledgements}
This work was supported by EPSRC (SD).

\appendix
\section{Measurement in the dual basis for KCMQ}
\label{DualBasisMeas}   

Here, we explain how to perform a dual basis ($ \ket{\pm} = (\ket{0} \pm \ket{1} )/\sqrt{2} $) measurement for KCMQ. $\ket{0}$ corresponds to the state peaked about $\Theta \equiv \varphi_2 - \varphi_1 = 0$, while $\ket{1}$ corresponds to the state peaked about $\Theta = \pi$. 
If $\ket{0,1}$ are ideal `spikes' at $\Theta = 0,\pi$, then $\ket{+,-}$ are superpositions of either even or odd number states, respectively. $\ket{+,-}$ can thus be thought of as even or odd parity states, respectively. A dual basis measurement therefore corresponds to a parity measurement of $n$ (the number operator, which is conjugate to $\Theta$). 
  
Attaching a charge measurement device - say a charge qubit (`Cooper-pair box') - to the terminals of the qubit is no good for measuring the parity of $n$. This is because charge qubits respond with unit period i.e. they have the same response to $n$ and $n+1$ offset or gate charge (in dimensionless units). This motivates the following observation by Kitaev: If the wire connecting terminals $1$ and $3$ is cut, then $2\varphi_2 - \varphi_3 - \varphi_1 =0$ (mod $2\pi$), from which it follows that the phase drop across the terminals becomes $\varphi_3 - \varphi_1 =  2( \varphi_2 - \varphi_1 ) \equiv 2 \Theta$ . The operator conjugate to the phase difference $2 \Theta$ is $n/2$. If we now connect the two terminals across a charge qubit, the charge qubit can now tell the difference between even and odd $n$ i.e. it can tell whether $n/2$ is an integer or half-odd integer.

\begin{figure}[t]
\begin{minipage}{\columnwidth}
\begin{center}
\resizebox{.5\columnwidth}{!}{\rotatebox{0}{\includegraphics{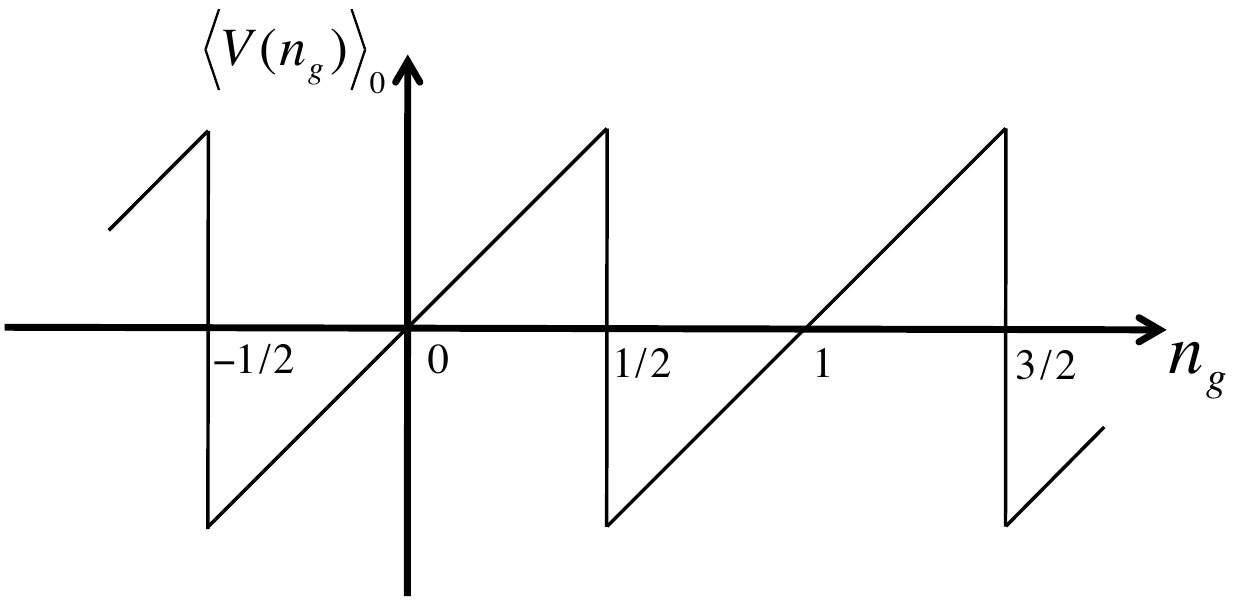}}}
\end{center}
\vspace*{-0.5cm} \caption{The ground state expectation value of the voltage as a function of $n_g$.} \label{sawtooth}
\end{minipage}
\end{figure}

The Hamiltonian for a charge qubit is

\begin{equation}
\mathcal{H} = 4E_C (N - n_g)^2 -E_J \cos \phi ,
\end{equation}  
where $[\phi , N] = i$, $E_C$ and $E_J$ are the charging and Josephson energies, respectively, and $n_g$ is the dimensionless external bias applied to the charge qubit. By attaching terminals $1$ and $3$ across the charge qubit, the bias becomes $n_g=n/2$. The voltage is given by $V = \frac{1}{2e} \frac{ \partial \mathcal{H} }{ \partial n_g}$, which leads to (using the Hellmann-Feynman theorem) the ground state expectation value of the voltage:

\begin{equation}
\langle V(n_g) \rangle_0 = \frac{1}{2e} \frac{\partial E_0}{\partial n_g} .
\end{equation}
This expectation value is periodic in $n_g$ with unit period, and resembles a `sawtooth', see Fig.~\ref{sawtooth}.  

Adding a further bias of about $1/4$ leads to a total bias of $n_g \approx 1/4 + n/2$. A voltage measurement of the charge qubit subsequently needs to be performed. If the measured voltage is positive, then $n$ is even, whereas if it is negative, then $n$ is odd. Since we are considering an elementary charge qubit operating in the regime where $E_C$ dominates $E_J$, its eigenstates are effectively the charge eigenstates (except near the `sweet spots' where $n_g$ is a half-odd integer). As a result, the measured voltage would be to a very good approximation equal to $\langle V \rangle_0$, and so the measurement is essentially one-shot. This scheme thus realises a parity measurement of $n$.       

\section{A vortex DC SQUID} 
\label{VortexDCSQUID}

\begin{figure}[t]
\begin{minipage}{\columnwidth}
\begin{center}
\resizebox{.37\columnwidth}{!}{\rotatebox{0}{\includegraphics{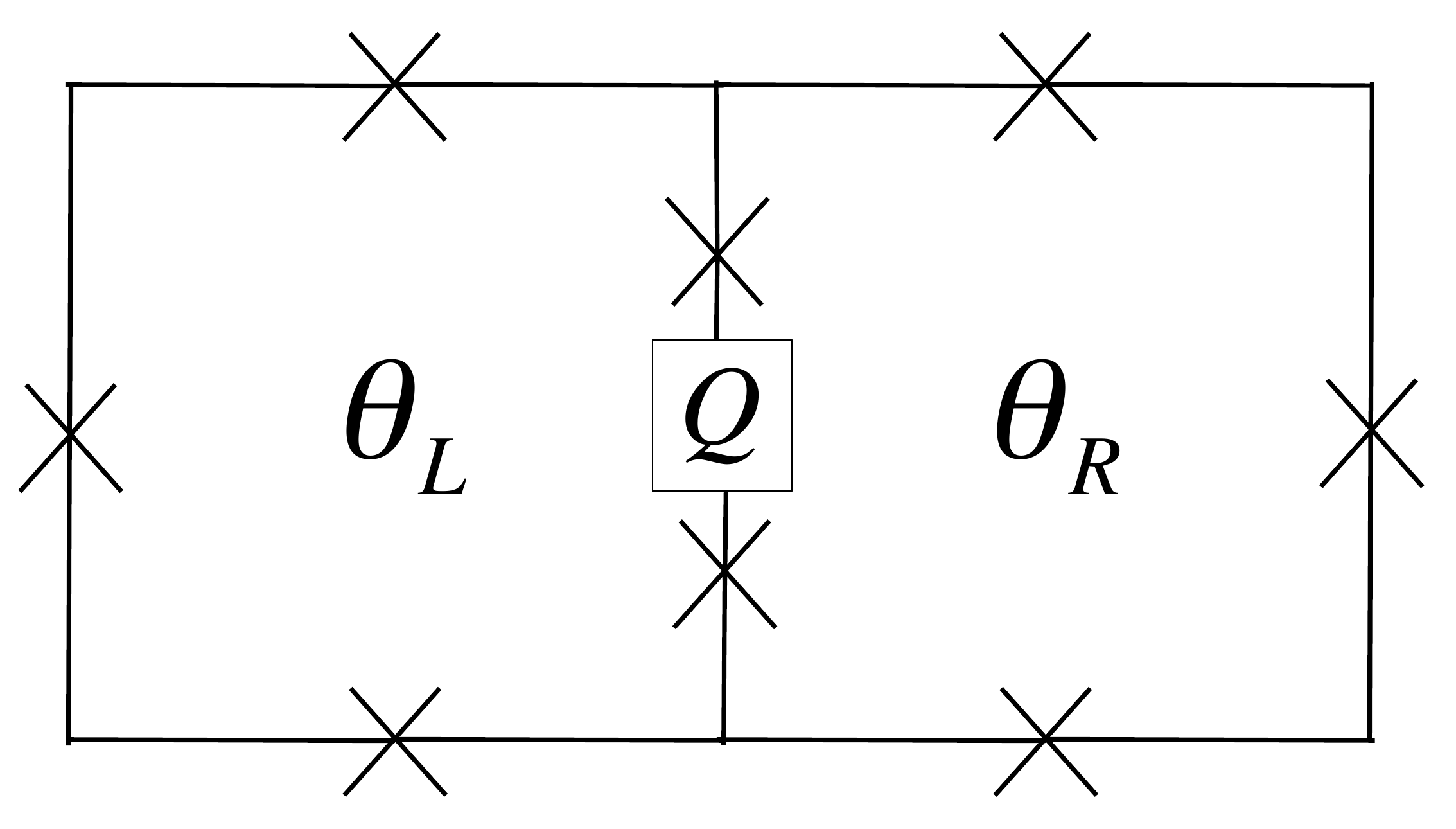}}}
\end{center}
\vspace*{-0.5cm} \caption{A vortex DC SQUID.} \label{DualSquid}
\end{minipage}
\end{figure}

To implement the GKP code, we require a switch. In KCMQ, this is simply a conventional DC SQUID \cite{Tinkham}. A vortex DC SQUID is sketched as follows (see Fig.~\ref{DualSquid}). There are two adjacent vortex sites, with phases $\theta_{L/R}$, and tunnelling between them can take place across two junctions which both have charging energy $E_C$. Charge $Q$ sitting on the island between the two junctions can control tunnelling, in the same way that external flux piercing a conventional DC SQUID controls charge tunnelling via parallel paths.
   
The two gauge invariant phase differences are

\begin{equation}
\theta_1 = \theta_R - \theta_L - \frac{\Phi_0}{\hbar} \int_L^R d \vec{r} \cdot \vec{A}_Q \; ,  \; \;  \theta_2 = \theta_L - \theta_R - \frac{\Phi_0}{\hbar} \int_R^L d \vec{r} \cdot \vec{A}_Q,
\end{equation}
where $\vec{A}_Q$ is the electric charge vector potential \cite{vanWees} satisfying $\oint d \vec{r} \cdot \vec{A}_Q = Q$. The energy of this component has the form expected of a DC SQUID, namely

\begin{equation}
E = -E_C \left( \cos \theta_1 + \cos \theta_2 \right) = -E_C \cos \left( \frac{\pi Q}{2e} \right) \cos \left( \theta_R - \theta_L \right).
\label{DualSQUIDEnergy}   
\end{equation}

\section{A vortex oscillator}
\label{VortexOsc}

\begin{figure}[t]
\begin{minipage}{\columnwidth}
\begin{center}
\resizebox{.37\columnwidth}{!}{\rotatebox{0}{\includegraphics{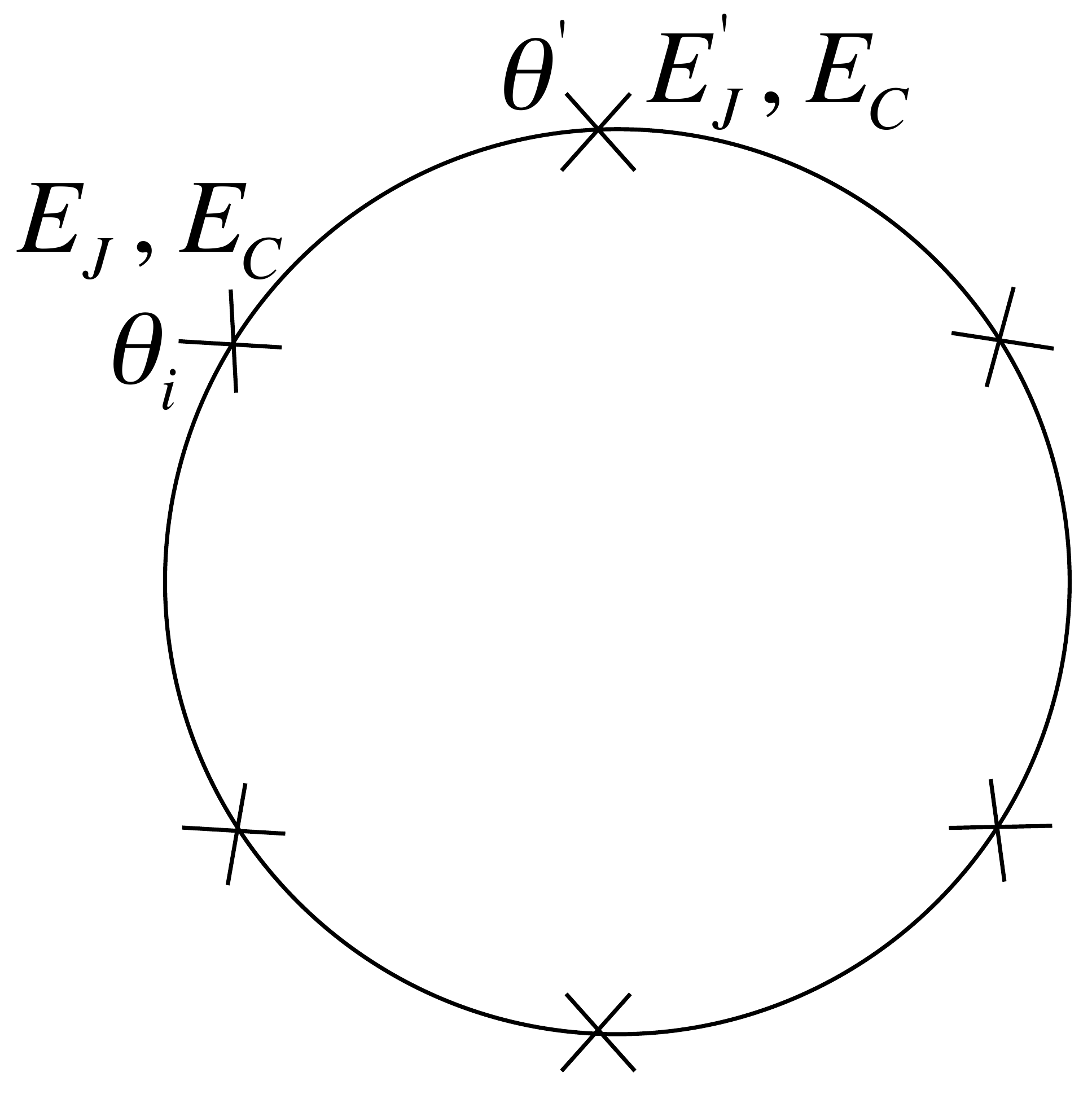}}}
\end{center}
\vspace*{-0.5cm} \caption{A ring of Josephson junctions.} \label{RingOfJJs}
\end{minipage}
\end{figure}

Let us first consider a simple toy model to establish the kind of phenomenology we would like to see in a vortex oscillator. Consider a ring of Josephson junctions, as shown in Fig.~\ref{RingOfJJs}. The junction at the `top' is characterised by junction energy $E_J'$ and charging energy $E_C$, while all other junctions are characterised by a junction energy $E_J$ and $E_C$. Let us assume that $E_J >> E_C >> E_J'$, which means that the top junction effectively looks like a capacitor, while all the others effectively look like inductors or semi-classical junctions. Denoting the number of charges on the top junction as $n'$ and the phase drop across it as $\theta '$, while the phases of the other junctions is $\theta_i$, the Hamiltonian of the ring is effectively

\begin{equation}
\mathcal{H} \approx 4 E_C n'^2 - \underbrace{ E_J \sum_i (\cos \theta_i -1) }_{f(-\theta')} .
\label{H_ring}
\end{equation}  
The fact that the Josephson energy term in (\ref{H_ring}) can be represented as $f(-\theta')$ follows from the loop constraint $\theta' + \sum_i \theta_i = 0 \rightarrow \sum_i \theta_i = -\theta'$, and the fact that the energy of a chain of junctions in the superfluid phase is a $2\pi$ periodic function ($f$) of the phase drop across the chain. Expanding about the minimum ($\theta' = 0$) where $f$ is quadratic, we can write $f(-\theta') \sim \theta'^2$, and so the system looks like a harmonic oscillator. 

Before we discuss our construction of a vortex oscillator, let us consider the electrostatics of vortices. It is well known that when vortices in an isotropic array have a separation which is much greater than the lattice constant, they can be considered to be point charges which interact via a two-dimensional (2D) Coulomb interaction. In a 2D world, the electric field due to a point charge $Q$ is simply $E = \frac{Q}{2\pi \epsilon_0 r}$, which can be derived from Gauss's law in two dimensions. As a result, the energy stored in a $\pm Q$ charge configuration separated by a distance $r$, relative to the energy stored at a distance $a$, is 

\begin{equation}
U = \frac{Q^2}{2\pi \epsilon_0} \ln \left(\frac{r}{a}\right).
\label{CoulombEnergy}
\end{equation}  
The vortex-vortex interaction term in an isotropic array with Josephson energy $E_J$ is $\pi E_J V_i I'_{ij} V_j$ (see (\ref{Sccg})), which motivated by (\ref{CoulombEnergy}) can be rewritten as $\frac{1}{2\pi \epsilon_0} 2 \pi^2 \epsilon_0 E_J V_i I'_{ij} V_j$. From this, we identify the unit of vortex charge to be $m \equiv \sqrt{2 \pi^2 \epsilon_0 E_J}$.

\begin{figure}[t]
\begin{minipage}{\columnwidth}
\begin{center}
\resizebox{.2\columnwidth}{!}{\rotatebox{0}{\includegraphics{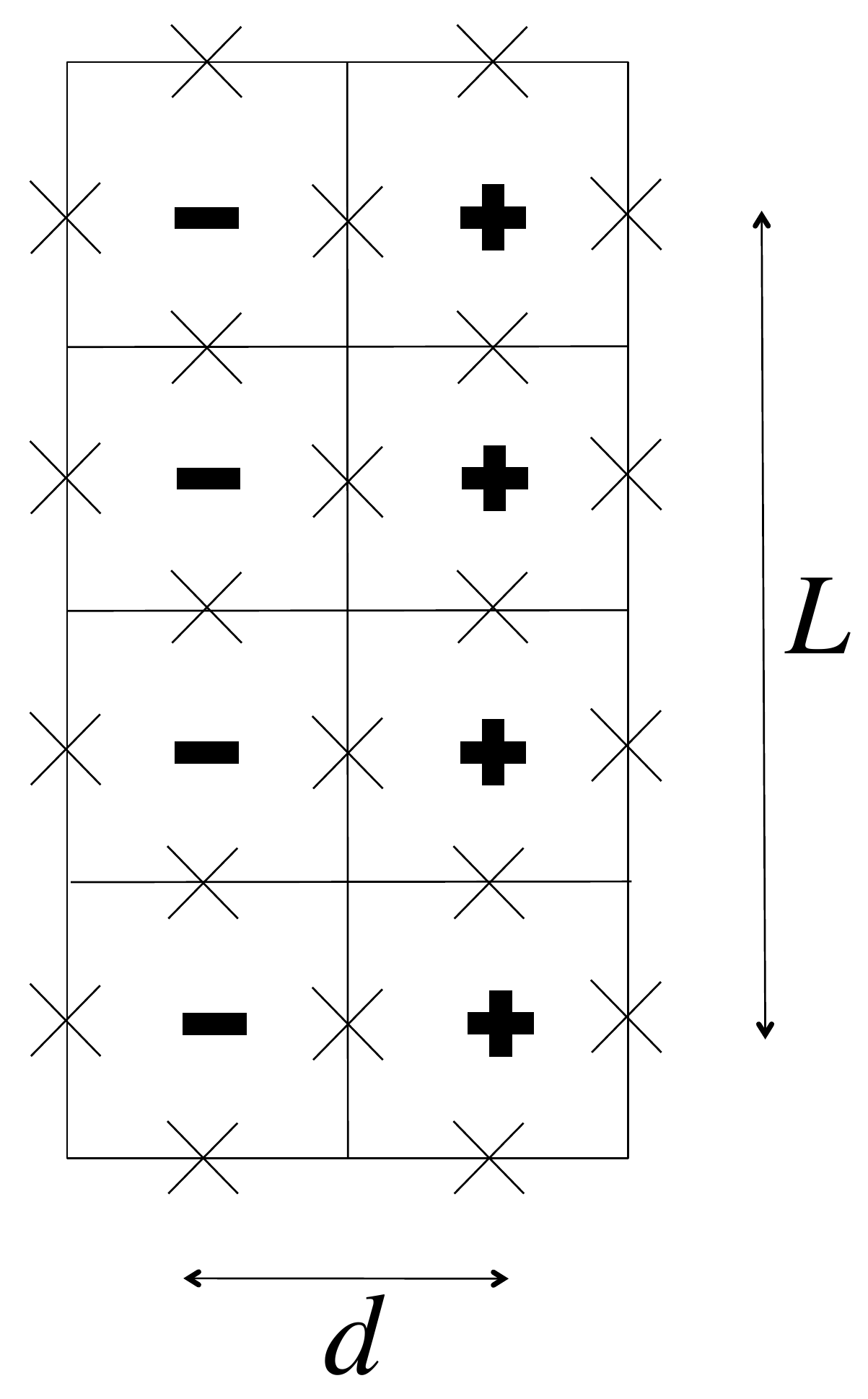}}}
\end{center}
\vspace*{-0.5cm} \caption{A 2D vortex capacitor.} \label{Cap}
\end{minipage}
\end{figure}

Treating vortices as Coulombic charges, one can perform a simple calculation of the energy stored in a 2D `vortex capacitor', along the lines of a calculation for electric charges in three dimensions. For a capacitor storing a $\pm Q$ charge configuration, the energy still has the form $Q^2 / 2C_{\rm{c}}$, with the capacitance being $C_{\rm{c}} = \frac{\epsilon_0 L}{d}$ (the subscript `c' in $C_{\rm{c}}$ denotes that $C_{\rm{c}}$ is the capacitance in conventional units). Here, $L$ is the length of the `plates' (in units of the dual lattice spacing) and $d=1$ is the distance between the `plates', as shown in Fig.~\ref{Cap}. The energy stored is thus 

\begin{equation} 
\frac{Q^2}{2C_{\rm{c}}} = \frac{[(L+1)m]^2}{2C_{\rm{c}}} = \pi^2 E_J \frac{(L+1)^2}{L}  .  
\end{equation}  

However, vortices only behave as Coulombic charges when the separation between them becomes large. One ought to check whether the simple $Q^2 / 2C_{\rm{c}}$ energy of the setup in Fig.~\ref{Cap} actually matches up with an exact calculation of the energy stored using $\pi E_J V_i I'_{ij} V_j$. It turns out that for small $L$, there is a large mismatch between the energies calculated via the two methods. However, the ratio of the energies calculated via the `$Q^2 / 2C_{\rm{c}}$' way and `the exact way' tends to unity as $L$ increases: for $L=20$ this ratio is approximately 1.2. Therefore, for large enough $L$, the energy of the configuration shown in Fig.~\ref{Cap} should be approximately given by $Q^2 / 2C_{\rm{c}}$.  

\begin{figure}[t]
\begin{minipage}{\columnwidth}
\begin{center}
\resizebox{.5\columnwidth}{!}{\rotatebox{0}{\includegraphics{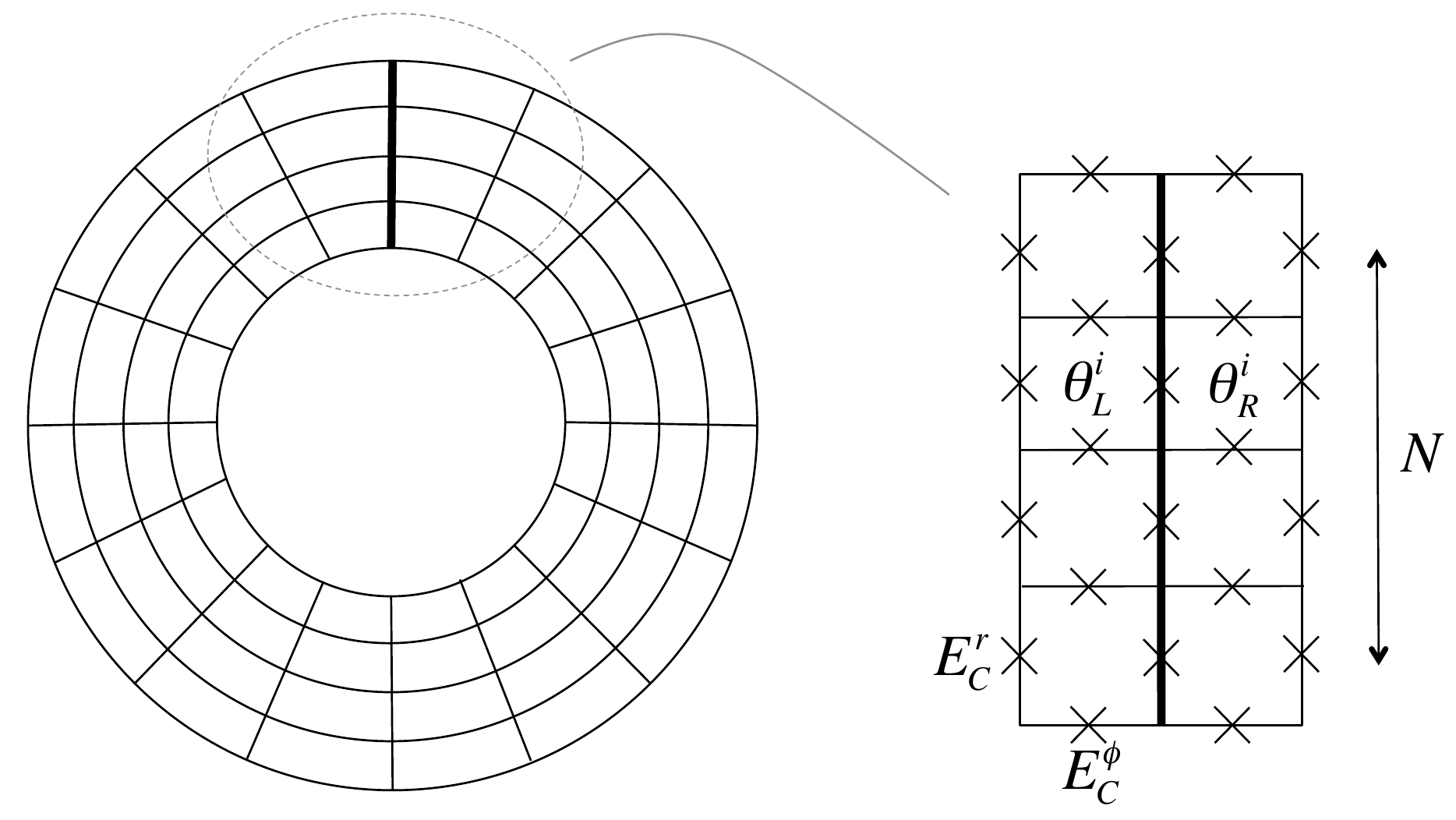}}}
\end{center}
\vspace*{-0.5cm} \caption{The array for a vortex oscillator (left) and `the capacitive bit' of the oscillator (right).} \label{VortexOscillator}
\end{minipage}
\end{figure}

Let us now discuss how a vortex oscillator may be realized. The array for such an oscillator is sketched in Fig.~\ref{VortexOscillator}. We assume that the Josephson energies $E_J$ are isotropic, and chosen (in conjunction with the charging energies to be discussed) such that free vortices and antivortices can exist in the array. Now, we take the charging energies of `the capacitive bit' (the junctions along the thick black line in Fig.~\ref{VortexOscillator}) and of junctions in the angular direction (except for the angular junctions right next to the thick black line, which have energy $E_C^{\phi}$) as being negligible. As a result, we can neglect vortex tunnelling across these junctions. If vortices and antivortices can `pile up' on either side of the thick black line, which behaves as a barrier for vortices, then this part of the array would resemble a `vortex capacitor', as in Fig.~\ref{Cap}. 

We would like each `track' of the array at a constant radius (starting from one side of the barrier and ending on the other side) to resemble a chain of dual junctions in the superfluid phase, as in the toy model sketched above. To achieve this, we need the charging energy in the radial direction to be as large as possible relative to $E_J$ such that a description in terms of vortices remains sensible: $E_C^{r} \gtrsim E_J$. Similarly, we also require $E_C^{\phi} \gtrsim E_J$, so that all the dual phases on one side of the barrier (right next to the barrier) become phase locked: $\theta_{L/R}^i \approx \theta_{L/R}^{i+1}$.  

Starting from the commutation relation $[\theta_i , V_j] = i\delta_{ij}$, we can make the change of variables

\begin{equation}
[ \underbrace{\theta^i_R - \theta^i_L }_{\gamma^i} , \underbrace{\frac{1}{2}(V^i_R - V^i_L) }_{Q^i /m} ] = i,
\end{equation}
where $m$ is the unit of vortex charge defined above. With the above conditions in place, the Hamiltonian describing the array is approximately

\begin{equation}
\mathcal{H} \approx \frac{1}{2C_{\rm{c}}} Q_{\rm{Tot}}^2 + \frac{E_C^r}{2M} \sum_{i=1}^{N+1} {\gamma^i}^2 + E_C^{\phi} \sum_{i=1}^{N} \sum_{X=L,R} [1 - \cos (\theta_X^{i+1} - \theta_X^{i} ) ].
\label{H_VO}  
\end{equation}       
In (\ref{H_VO}), $C_{\rm{c}} = \epsilon_0 N$, $Q_{\rm{Tot}} = \sum_{i=1}^{N+1} Q^i$ and $M$ is the number of junctions (with hopping amplitude $E_C^{r}$) in each track. The second term reflects the fact that each track resembles a chain of junctions in the superfluid phase (see the discussion of the toy model above and the chain of superfluid junctions in Sec.\ref{KCMQ}). If $E_C^{\phi}$ is strong enough such that $\theta_{L/R}^i \approx \theta_{L/R}^{i+1}$, then all the phase drops across the barrier are nearly the same, say $\gamma^i \approx \gamma$. As a result,    

\begin{equation}
\mathcal{H} \approx \frac{1}{2C_{\rm{c}}} Q_{\rm{Tot}}^2 + \frac{E_C^r (N+1)}{2M} \gamma^2 ,
\label{H_VO1}
\end{equation}
with $\gamma$ and $Q_{\rm{Tot}}$ being conjugate variables since

\begin{equation}
\left[ \underbrace{\frac{1}{N+1}  \sum_{i=1}^{N+1} \gamma^i }_{\gamma} ,  \frac{1}{m} \underbrace{ \sum_{i=1}^{N+1} Q^i }_{Q_{\rm{Tot}}} \right] = i.
\end{equation}
The system we have described is thus effectively a vortex harmonic oscillator.

We can rewrite (\ref{H_VO1}) in the following form

\begin{equation}
\mathcal{H} \approx \frac{q_{\rm{Tot}}^2}{2C}  + \frac{\gamma^2}{2L},
\label{H_VO2} 
\end{equation}
where $q_{\rm{Tot}} = Q_{\rm{Tot}}/m$, and $C = C_{\rm{c}}/m^2$ and $L=\frac{M}{(N+1) E_C^r}$ are the capacitance and inductance of the oscillator, respectively. Both $L$ and $C$ have dimensions of inverse energy. For the gate $R(\pi/4)$ to be fault-tolerant, the impedance of the oscillator, $\sqrt{L/C}$, should be large. We can express $L/C$ in terms of parameters of the array:

\begin{equation}
\frac{L}{C} = \frac{M}{(N+1) E_C^r} \cdot \frac{m^2}{C_{\rm{c}}} = \frac{M}{(N+1) E_C^r} \cdot \frac{2\pi^2 \epsilon_0 E_J}{\epsilon_0 N} = \frac{2\pi^2 M E_J}{N(N+1) E_C^r}.    
\end{equation}
Since $M$ scales with the radius of the `hole' of the array, by making this radius arbitrarily large compared to the capacitor plate length, the impedance can be made arbitrarily large.

\end{document}